\documentclass[12pt,a4paper]{article} 
\pdfoutput=1

\usepackage{jheppub}

\usepackage{graphicx}
\usepackage{amsmath}
\usepackage{amssymb}
\usepackage{marginnote}
\usepackage{relsize}
\usepackage[bottom]{footmisc}

\def\beq{\begin{equation}}
\def\eeq{\end{equation}}
\def\beqa{\begin{eqnarray}}
\def\eeqa{\end{eqnarray}}

\title{Instability of ${\mathcal N}=2$ gauge theory in compact space with an
  isospin chemical potential}

\author{Suphakorn Chunlen, Kasper Peeters, Pichet Vanichchapongjaroen\\
  and Marija Zamaklar}

\affiliation{Department of Mathematical Sciences,\\
Durham University,\\
South Road,\\
Durham DH1 3LE,\\
United Kingdom.}

\emailAdd{suphakorn.chunlen@durham.ac.uk} 
\emailAdd{kasper.peeters@durham.ac.uk} 
\emailAdd{pichet.vanichchapongjaroen@durham.ac.uk}
\emailAdd{marija.zamaklar@durham.ac.uk}

\preprint{DCPT-12/41}

\abstract{We investigate ${\mathcal N}=2$ super-Yang-Mills theory on a
  three sphere in the presence of an isospin chemical potential at
  strong coupling using the AdS/CFT correspondence. This system
  exhibits an instability for sufficiently large values of the
  chemical potential. In contrast to other related models, the first
  excitation to condense is not a vector meson but rather a scalar
  charged under the global $SO(4)$ symmetry group. Furthermore, the
  spectrum of fluctuations exhibits an interesting cross-over
  behaviour as a function of the dimensionless temperature. We
  construct the new ground state of the non-linear theory.}

\begin{document}
\maketitle

\section{Introduction and summary}

Various methods and models indicate that in QCD, the presence of a
sufficiently large isospin chemical potential leads to condensation of
mesonic particles. Such new ground states may have interesting
properties, such as non-isotropy or even non-homogeneity.
Intuitively, condensation will take place when the chemical potential
is of the order of the mass of the lightest meson in the system. In
QCD, this would thus imply that a modest chemical potential leads to the
formation of a pion condensate, and increasing the potential further
is then expected to trigger rho meson condensation and so forth. While
this intuitive picture is simple, computing such condensation effects
from first principles is hard, though is indeed suggested by
phenomenological models
\cite{Voskresensky:1997ub,Lenaghan:2001sd,Sannino:2002wp} as well as
studies that make use of the string/gauge theory correspondence
\cite{Aharony:2007uu,Ammon:2008fc,Ammon:2009fe} (though not all of
these contain the analogue of a pion sector). Obtaining these results
from the lattice still remains a challenge.

When additional scales are present in the problem, it is conceivable
that more interesting or non-intuitive things happen. One particular
situation that we will be interested in here is the effect of
\emph{finite volume}. In the context of the string/gauge theory
correspondence, there is a canonical example involving gauge theory in
a space with finite volume, the dual description of which involves
string theory on $\text{AdS}_5\times S^5$ in global
coordinates. According to the conjecture, this describes ${\mathcal N}=4$
super-Yang-Mills theory on the compact manifold living on the boundary
three-sphere (plus time). In its original form~\cite{mald2}, the
string/gauge theory correspondence links string theory on the Poincar\'e
patch of $\text{AdS}_5\times S^5$ with ${\mathcal N}=4$ super-Yang-Mills theory
on the non-compact $\mathbb{R}^{3,1}$. The Poincar\'e patch of the
string theory background arises as the near-horizon limit of the
supergravity solution for $N_c\rightarrow \infty$ coincident
D3-branes. This brane construction serves as the starting point for
the ``derivation'' of the correspondence. In contrast, global
$\text{AdS}_5\times S^5$ space does not arise as the near horizon
limit of any D-brane configuration. However, it is strongly believed
that the correspondence holds here as well, in the manner described
above. Generalisation to finite temperature is straightforward and
amounts to compactifying the time circle both on the gauge and gravity
sides, with the size of the thermal circle $S^1$ being inversely
proportional to the temperature of the system. When the
temperature is high enough, the system undergoes a Hawking-Page phase
transition and empty thermal $\text{AdS}$ space is replaced with a
Schwarzschild black hole with a spherical
horizon~\cite{Witten:1998zw}. The existence of this phase transition
is possible because of the presence of a dimensionless parameter (the
ratio of the radii of the thermal circle and of the three-sphere), which
is not present at infinite volume.

In order to study mesonic excitations in this model, one has to add
matter fields to it, or in holographic language, to add probe branes.
In order to stay as close as possible to the ${\mathcal N}\!=\!4$ model, one can
add D7-branes, which was done in global AdS
in~\cite{Karch:2006bv}. This modification of the system breaks
supersymmetry down to ${\mathcal N}\!=\!2$ and incorporates quarks in the
fundamental representation of the gauge group. If the bare quark mass
is chosen to vanish, the conformal invariance of the system is
preserved. The isometries of the internal three-sphere give rise to a
global $SO(4)$ R-symmetry, and for $N_f$ branes, one has in addition an
$SU(N_f)$ flavour symmetry. Chemical potentials can be introduced for
all of these, but in the present paper we will study the effect of a
chemical potential associated to the latter, and will call it an
`isospin potential' (see \cite{PremKumar:2011ag} for an analysis of
some of the effects of an R-symmetry potential).

% 
% 
% 
% While the dual meson spectrum obtained from string theory on the
% Poincar\'e patch contains a massless multiplet~\cite{Kruczenski:2003be}
% 

\medskip

In this compact gauge theory, we find two features which to our
knowledge have not been observed in the literature before. The first
is that the particles which condense first are \emph{not} vector
excitations (`rho mesons'), but rather scalar particles charged under
the global symmetry group.\footnote{This could perhaps have been
  expected from the analysis of the meson spectrum of
  \cite{Erdmenger:2010zm}, which finds that these $SO(4)$ charged
  scalar mesons are the lightest, but their work does not discuss the
  effect of a chemical potential, and the qualitative argument based
  on only a comparison of the mass may in any case not be the complete
  story.}
This is an effect which in fact survives the large radius limit at
fixed temperature, i.e.~the limit towards a black hole in the
Poincar\'e patch. The anisotropy of the condensate is thus in the
direction of the internal manifold, in contrast to what has been
assumed in earlier literature for this model
\cite{Ammon:2008fc,Erdmenger:2008yj,Ammon:2009fe}, and in contrast to
results in the Sakai-Sugimoto model~\cite{Aharony:2007uu}.

The second feature is that the thermal pole masses of some of the
mesons can cross as the dimensionless ratio is varied.\footnote{This
  crossing behaviour is somewhat reminiscent of the mixing of states
  observed in~\cite{Aharony:2007uu}.} This suggests that, as the
chemical potential is increased from zero, the particles which
condense first to form a new ground state may not always be the same,
but depend on this parameter. We indeed confirm this crossing
behaviour also for the condensate formation, by explicit computation
of the fluctuation spectrum, construction of the condensate solutions
and a computation of their free energies. In our model the crossing
does not involve the lightest particle, so the ground state remains a
condensate of $SO(4)$ charged scalars, but in other models the
situation may not be so simple, and it would be interesting to
investigate this further.

\bigskip

\section{Holography with a dual $S^3$}
\subsection{Brane embeddings in global $\text{AdS}_5\times S^5$ and $\text{AdS}_5$-Schwarzschild}

In this section we will briefly review some properties of the global
$\text{AdS}_5\times S^5$ and $\text{AdS}_5$-Schwarzschild spaces as well as various
embeddings of D7-probe branes in these
geometries~\cite{Karch:2006bv}. String theory in global $\text{AdS}_5\times
S^5$ space is believed to be dual to the ${\mathcal N}=4$ SYM theory at zero
temperature, which lives at the boundary of this space, which is an
$S^3\times \mathbb{R}$. Turning on the temperature makes the time
direction (both in the gauge and gravity sides) compact, with the
radius being inversely proportional to the gauge theory temperature.

The metric of the global $\text{AdS}_5$ spaces (both at zero and finite
temperature) is given by 
\begin{equation}
\label{AdSatT=0}
{\rm d}s^2=
   \left(1+\frac{r^2}{R^2}\right){\rm d}\tau^2 
  + \frac{{\rm d}r^2}{1+\frac{r^2}{R^2}} + r^2 {\rm d}\Omega^2_3 +
  R^2{\rm d}\Omega_5^2 \,,
\end{equation}
where the origin of the AdS space is at $r=0$, the boundary is at
$r\rightarrow \infty$, and $R$ is the $\text{AdS}$ radius. The
Euclideanised time direction $\tau$ is periodic with period
$R_\tau$. Since the theory is conformal, it implies that only the
ratio of the thermal circle $\tau$ and the size of the boundary sphere
on which dual theory is living, $R/R_\tau$, is physically observable,
i.e.~unchanged by the conformal symmetry of the theory. 

%Therefore, we see that the zero
%temperature limit is equivalent to the infinite volume limit in the dual
%theory.

As the temperature of the thermal $\text{AdS}$ space is increased, a
first order Hawking-Page phase transition takes place and global
AdS-Schwarzschild space replaces the thermal AdS space as the proper
ground state of the system~\cite{Witten:1998zw}.  The metric of the
global AdS-Schwarzschild black hole is given by
\begin{equation}
\label{AdSatTfinite}
{\rm d}s^2=
   - \left(1+\frac{r^2}{R^2}-\frac{M^2}{r^2}\right){\rm d}t^2
  +\frac{{\rm d}r^2}{1+\frac{r^2}{R^2}-\frac{M^2}{r^2}}+r^2{\rm
    d}\Omega^2_3+R^2 {\rm d}\Omega_5^2\,,
\end{equation}
where $M^2= 8 G_N m_{\text{bh}} /(3\pi)$ and $m_{\text{bh}}$ is the
mass of the black hole, while its temperature is given by
\begin{equation}
T=\frac{1}{4\pi}\left(\frac{2r_0}{R^2}+\frac{2M^2}{r_0^3}\right),
\end{equation}
where
\begin{equation}
r_0^2=R^2\left(\frac{-1+\sqrt{1+4M^2/R^2}}{2}\right) \, .
\end{equation}
The dimensionless ratio $R/R_\tau$ is proportional to $TR$, and we
will later on expand our results for large values of this parameter,
interpreting this limit as one at finite temperature and large
volume~\cite{Karch:2006bv,Karch:2009ph}, so that a comparison with
results in the Poincar\'e patch can be made.

Introducing D7-probe branes in this geometry corresponds, in the
holographic language, to adding flavour hypermultiplets to
$\mathcal{N}=4$ SYM on the sphere $S^3$. A study of various D-brane
probes, in particular D7-probe branes, in these geometries was
performed in \cite{Karch:2006bv,Karch:2009ph}. It was found that at
zero (and low) temperature, there are two possible D7-brane embeddings
in this dual geometry. The first type of embeddings are those in which
the D7-brane completely fills the $\text{AdS}_5$ space and wraps the $S^3 \in
S^5$ which is equatorial (i.e~it is a D-brane with vanishing extrinsic
curvature). The second type of embeddings are those in which the
D7-brane wraps a non-maximal $S^3\in S^5$ which shrinks along the
radial direction of $\text{AdS}_5$ and becomes zero before the brane reaches
the origin of $\text{AdS}_5$. The first series is dual to the $\mathcal N=2$
SYM theory with massless hypermultiplets, while for the second class
the hypermultiplet is massive, and its mass is related to the distance
at which the D7-brane ``stops'' before the origin of the
$\text{AdS}_5$. Interestingly, as the ``quark'' mass is varied a topology
changing phase transition occurs \cite{Karch:2006bv}. It was further
analysed in detail in \cite{Karch:2009ph} that this phase
transition is actually third order, unlike most of the phase
transition associated to probe branes in holographic duals, which are
usually first order.

In the high temperature phase, the situation is similar to that in
infinite volume. One finds that Lorentzian and black hole embeddings
exhaust all possibilities. As for infinite volume, these correspond to
D7-branes that stay outside the horizon or reach the black hole
horizon respectively.

In this paper we will be studying various D-brane embeddings in the
presence of an ``isospin'' chemical potential. For that purpose,
instead of using the coordinates \eqref{AdSatT=0}, it will be useful
to use another set of coordinates given by
\begin{equation}\label{eq:metric}
{\rm d}s^2=
   -\frac{u^2}{R^2}\left(1+\frac{R^2}{4u^2}\right)^2{\rm d}t^2
   + u^2\left(1-\frac{R^2}{4u^2}\right)^2{\rm d}\bar{\Omega}_3^2
   +\frac{R^2}{u^2}(du^2+u^2{\rm d}\Omega_5^2)\,,
\end{equation}
which is related to \eqref{AdSatT=0} via the coordinate change
\begin{equation}
u=\frac{1}{2}(r+\sqrt{r^2+R^2}) \, .
\end{equation}
In these coordinates the origin of the $\text{AdS}$ space is at $u=R/2$, while
the boundary is an $S^3$ at $u\rightarrow \infty$. We will also use
\begin{equation}
{\rm d}s^2 = 
   - \frac{1}{4z^2} \left(1 + z^2\right)^2 {\rm d}t^2 
   + \frac{R^2}{4z^2} \left(1-z^2 \right)^2 {\rm d}\bar{\Omega}_3^2  
   + R^2 \frac{{\rm d}z^2}{z^2} + R^2 {\rm d}\Omega^2 \, ,
\end{equation}
which is related to the previous coordinates by $z=R/(2u)$. Let us
also note that in the $z$ coordinate the origin of AdS space is at
$z=1$, while the boundary is at $z=0$.

Similarly for the system at finite temperature in addition to metric
\eqref{AdSatTfinite} we will also use $(u,t)$ coordinates,
\begin{equation}
\label{AdSfiniteV}
{\rm d}s^2=-\frac{2\rho_H^2}{u R^2} \frac{F(u)}{W(u)}{\rm d}t^2
  +2\frac{\rho_H^2}{u}W(u){\rm d}\bar{\Omega}_3^2
  +\frac{R^2}{4u^2 F(u)}{\rm d}u^2+R^2{\rm d}\Omega_5^2\,,
\end{equation}
where
\begin{equation}
F(u)=1-u^2, \quad \frac{\rho_H^4}{R^4} = \frac{1}{16} + \frac{M^2}{4 R^2}, \quad 
W(u)=1-\frac{uR^2}{4\rho_H^2} \, .
\end{equation}
Note here that the variable $u$ is dimensionless and ranges from $u
\in [0, 1]$, where the horizon is at $u=1$, and the boundary is at
$u=0$. The Hawking temperature of this black hole is
\begin{equation}
T = \frac{\sqrt{2}\rho_H}{\pi R^2 \sqrt{1-\frac{R^2}{4\rho_H^2}}}\,.
\end{equation}

For numerical investigation it turns out that another form of the
metric at finite $T$ will also be useful.\footnote{This is related to
  the fact that the solutions at vanishing bare quark mass can be
  related to Heun functions in these coordinates.} If we change
coordinates as
\begin{equation}
v=\frac{1-u}{1-\frac{uR^2}{4\rho_H^2}} \quad \quad
\Lambda=\frac{8\rho_H^2}{4\rho_H^2+R^2} \, ,
\end{equation}
then the metric becomes
\begin{equation}
\label{coordinatev}
{\rm d}s^2=-\frac{v(\Lambda-v)}{(1-v)(2-\Lambda)}{\rm d}t^2
   +\frac{\Lambda-1}{(2-\Lambda)(1-v)}R^2{\rm d}\bar{\Omega}_3^2
   +\frac{\Lambda-1}{4(1-v)^2v(\Lambda-v)}R^2{\rm d}v^2+R^2{\rm d}\Omega_5^2\,,
\end{equation}
where 
\begin{equation}
{\rm d}\Omega^2_3=\frac{1}{4}({\rm d}\alpha^2+{\rm d}\beta^2+{\rm
  d}\gamma^2+2\cos\beta {\rm d}\alpha d\gamma).
\end{equation}
In these coordinates $v=0$ is the horizon while $v=1$ is the boundary,
and $\Lambda$ is function of the temperature given by
\begin{equation}
\Lambda =\frac{4 (\pi TR)^2}{3(\pi TR)^2-\sqrt{(\pi TR)^2 \left(-2+(\pi TR)^2\right)}}.
\end{equation}
% Finally, let us mention that in the limit of large radius of the
% boundary sphere (i.e.~taking the limit $TR \rightarrow \infty$), the
% metric and physics in global $\text{AdS}$ space reduces to that in the
% Poincar\'e patch. We will take this limit repeatedly throughout the
% paper in order to check our results against known results in the
% Poincar\'e patch.

\subsection{Chemical potentials and homogeneous solutions}

In this section we will construct solutions which correspond to adding
a chemical potential associated to the global $SU(2)$ flavour
symmetry. This symmetry originates from the fact that we are
considering two coinciding D7-probe branes. In addition to the $SU(2)$
symmetry, our systems also exhibits another global $SO(4)$ symmetry
associated to the residual global isometry of the system of probes. In
principle one could also consider switching on a chemical potential
which is associated to this symmetry group, as it was done in
e.g.~\cite{PremKumar:2011ag}. However, our prime interest will be the
physically more relevant $SU(2)$ group, which has direct analogue with
the $SU(2)$ flavour symmetry group in the Sakai-Sugimoto
model~\cite{Aharony:2007uu}.

In order to turn on a chemical potential corresponding to this global
$SU(2)$ ``isospin'' symmetry, let us consider two coincident D7-branes,
which for simplicity have equatorial embedding $\theta = 0$, so that
the induced metric on the D7-branes is
\begin{equation}
\label{induceds}
{\rm d}s^2=
    -\frac{u^2}{R^2}\left(1+\frac{R^2}{4u^2}\right)^2{\rm d}t^2
    +\frac{R^2}{u^2} {\rm d}u^2  +
    u^2\left(1-\frac{R^2}{4u^2}\right)^2{\rm d}\bar{\Omega}_3^2+ R^2
    {\rm d}\Omega_3^2\,.
\end{equation}
In other words, the D7-probes fill out the full $\text{AdS}_5$ space, as well
as a maximal $S^3 \in S^5$. We should note that there are two $S^3$
factors present on the world-volume of the brane, one $\bar{\Omega}_3$
which is dual to the boundary $S^3$ and another one $S^3$ in $S^5$,
which is part of the global symmetry group.

As explained in the previous section, in order to turn on a chemical
potential we need to turn on the $A_0$ component of the gauge field,
such that it satisfies the boundary condition
\begin{equation}
\label{boundaryA0}
A_0(x,z\rightarrow 0) \rightarrow \mu_I \tau^3\, . 
\end{equation}
As a first guess for finding the ground state of the system in the
presence of this chemical potential, we consider the homogeneous ansatz
\begin{equation}
A=A_0^{(3)}(u)\tau^3\, {\rm d}t\, , 
\end{equation}
so that the DBI action becomes
\begin{equation}
S = -T_{D7}8\pi^4R^3  \int\! {\rm d}z\sqrt{-g(z)}\bigg[1+\frac{\pi^2R^4}{2\lambda}\left(\partial_zA_0^{(3)}(z)\right)^2g^{tt}(z)g^{zz}(z)\bigg]^{1/2} \, .
\end{equation}
The equation of motion for the field $A_0(z)$ can be integrated once, yielding
\begin{equation}
\partial_z(A_0^{(3)}(z)R)=\frac{4cz(1+z^2)}{\sqrt{(1-z^2)^6+32\frac{c^2\pi^2}{\lambda}z^6}} \, , 
\end{equation}
where $c$ is an integration constant.  We are looking for a physical
configuration which is smooth and differentiable
everywhere. Specifically, we require that the field $A_0$ and its
derivatives are smooth at the origin of AdS space. However, by
expanding the right hand side of the above equation near the origin of
AdS space one sees that the radial derivative of the $A_0$ field is
non-vanishing.  In other words, it is not possible to obtain any
nontrivial (different from a constant solution) homogeneous solution
which is smooth at the origin. This observation persists also for the
Yang-Mills truncation of the DBI action, and it also holds in the
Poincar\'e limit (see \cite{Kobayashi:2006sb}). Hence as the starting
configuration for our fluctuation analysis we will use a homogeneous
solution with non-zero isospin potential, given by
\begin{equation}
\label{simplesoln}
A_0 = a_0 \tau ^3\,,  \qquad a_0 =  \text{const}. \, .
\end{equation}
This solution implies that at zero temperature, the homogeneous
background does not lead to generation of an isospin density. 

Above the Hawking-Page transition, the situation is similar. The homogeneous
ansatz yields a first order differential equation,
\begin{equation}
\partial_uA_0^{(3)}(u)\left(1-\frac{u R^2}{4\rho_H^2}\right)^2= a_0 \,,
\end{equation}
where $a_0$ is an integration constant. This is solved by
\begin{equation}
A_0^{(3)}(u)=\frac{\mu_I(1-u)}{1-\frac{u R^2}{4\rho_H^2}},
\end{equation}
where we have imposed the boundary condition that $A_0(u\rightarrow
0)=\mu_I$ at the boundary, and also required vanishing of $A_0$ at the
horizon of the black hole in $\text{AdS}$ space. In contrast to the low
temperature situation, we see that there is now a non-vanishing isospin
density present, even for this homogeneous system.
\bigskip

\section{Perturbative analysis of the homogeneous vacuum at $T=0$}

While the homogeneous and isotropic solution which was discussed in
the previous section is a legitimate solution to the equations of
motion, we expect that for large enough values of the chemical
potential this configuration will become unstable and ``decay'' into
another, presumably non-homogeneous or non-isotropic ground state. Our
expectations are based on a similar analysis which was previously
performed for the Sakai-Sugimoto model in \cite{Aharony:2007uu}. A major
difference, however, with respect to the analysis of that paper is
that we are now dealing with a field theory on a \emph{compact}
space. In the present section we will discuss the perturbative
stability analysis at $T=0$, which from a technical perspective
largely follows the meson spectrum analysis
of~\cite{Erdmenger:2010zm}; we recall some elements of that
construction for completeness and in order to be able to compare with
the finite temperature analysis which is to follow in section~\ref{perturbfinite}.

\subsection{Scalar fluctuations at zero temperature}

We will start the perturbative analysis of the homogeneous solution
\eqref{simplesoln} by considering scalar perturbations. By scalars we
here mean scalars in the dual theory that are also scalars from the
point of view of the D7-probe, i.e.~gauge theory scalar fields which
are uncharged under the $SO(4)$. Some of the scalars in the dual gauge
theory originate from the components of the gauge field on the
D7-probe and will be analysed in the next section.  Our starting point
is the flat (i.e.~maximal) D7-probe embedding with the world-volume
field \eqref{simplesoln} turned on. The induced metric on the
world-volume was written in \eqref{induceds}. Since the D7-probe brane
is filling out the full $\text{AdS}_5$ space and wrapping a maximal
$S^3$ in $S^5$, there are only two transverse scalars to the brane
world-volume, and they are within the $S^5$. To see which scalars
these are, let us write the metric on $S^5$ as
\begin{equation}
{\rm d}s_5^2 = R^2({\rm d}\theta^2+\sin^2\theta\, {\rm
  d}\phi^2+\cos^2\theta\, {\rm d}\Omega_3^2) \, .
\end{equation}
Instead of using the $(\theta,\phi)$ coordinates it will be more convenient to 
introduce coordinates
\begin{equation}
w_1=R\sin\theta\cos\phi \, \quad w_2=R\sin\theta\sin\phi ,
\end{equation}
so that the metric on $S^5$ becomes
\begin{equation}
\label{S5metrics}
{\rm d}s_5^2 = \left(1-\frac{w_1^2+w_2^2}{R^2} \right) {\rm
  d}\Omega_3^2 + {\rm d}W^2(w_1,w_2)\,.
\end{equation}
Here ${\rm d}W^2$ is a complicated expression in terms of $w_1, w_2$, which
however significantly simplifies for $w_1=0, w_2=w$, as it then becomes
\begin{equation}
{\rm d}W^2(w_1,w_2)\to\frac{{\rm d}w^2}{1-\frac{w^2}{R^2}}\,.
\end{equation}
It can easily be checked from the equations of motion that it is
indeed consistent to set one of the $w_1, w_2$ to zero.\footnote{The
  fluctuation in the other direction leads to the same spectrum, so we
  will not comment on it any further (though we should emphasise that
  this is a property of the equatorial embedding not shared by
  non-zero bare quark mass embeddings).}
%Also, since we
%expect that the new ground state will not break the global $SO(4)$
%symmetry (corresponding to the isometry group of $S^3 \in S^5$) when
%considering flucutations we focus only on the $SO(4)$
%singlets. 
We expect that the first instability will appear already for the
lowest lying (S-wave) mode on the dual gauge theory sphere
$\bar{S}^3$, which is also a singlet on $S^3 \in S^5$.  Therefore,
when looking for the instabilities in the system, we will look at the
fluctuation which is a function of only the time and $u$-coordinates.
Let us define the fluctuation variable as
\begin{equation}
\label{e:Psitu_def}
\Psi(t,u)= \delta w_2(t,u)  \, .
\end{equation}
The induced metric on the D7-brane becomes  
\begin{multline}
{\rm d}s^2=-\frac{u^2}{R^2}\left(1+\frac{R^2}{4u^2}\right)^2{\rm d}t^2+u^2\left(1-\frac{R^2}{4u^2}\right)^2{\rm d}\bar{\Omega}_3^2 \\[1ex]
+\frac{R^2}{u^2}{\rm d}u^2+R^2\left(1-\frac{\Psi(t,u)^2}{R^2}\right){\rm d}\Omega^2_3 +\left(\frac{\partial \Psi(t,u)}{\partial t}{\rm d}t+\frac{\partial \Psi(t,u)}{\partial u}{\rm d}u\right)^2\,.
\end{multline}
We next need to write down the action for the scalar fluctuation to
leading order in~$\alpha'$. A subtle point here is that all scalars
are in the adjoint representation of the $SU(2)$ group on the
world-volume of two D7-branes. The approach we adopt here to write the
action, is to first treat all scalars as \emph{abelian} and derive the
action for the fluctuation by linearising the DBI action. In the last
step we then promote all fields to be in the adjoint representation by
introducing an overall trace in front of the action (for more on this
and other approaches, see e.g.~\cite{Myers:1999ps,Ammon:2009fe}).

Following these steps we end up with the action governing the scalar
fluctuations,
\begin{multline}
S=-T_{D7}\frac{4\pi^6 R^4}{\lambda}\int\!
{\rm d}u{\rm d}t\,u^3\left(1+\frac{R^2}{4u^2}\right)\left(1-\frac{R^2}{4u^2}\right)^3
\\[1ex]
\times\left[-\frac{R^2}{2}\left(\frac{4u}{4u^2+R^2}\right)^2 D_t \Psi^{(a)} D_t \Psi^{(a)}+\frac{u^2}{2R^2}\partial_u \Psi^{(a)} \partial_u \Psi^{(a)}-\frac{3}{2}\frac{\Psi^{(a)}\Psi^{(a)}}{R^2}\right] \, .
\end{multline}
Since we expect that an instability will appear in the gauge direction
orthogonal to the background field $A_0$, we make the ansatz for the
fluctuation field to be
\begin{equation}
\Psi(t,u) =  e^{-i\omega t} \left(\Psi^{(1)}_\omega(u)\tau^1+\Psi^{(2)}_\omega(u)\tau^2\right) \,,
\end{equation}
where we have focused on one Fourier mode.

The equations of motion for the components $\Psi^{(1)}_\omega$ and
$\Psi_\omega^{(2)}$ are coupled, but can be decoupled by changing
variables as
\begin{equation}
\Psi_\omega^{(\pm)}(u) =  \Psi^{(1)}_\omega(u) \pm i \Psi^{(2)}_\omega(u) \, .
\end{equation}
The equations of motion for the components $\Psi^{(\pm)}$ are 
\begin{equation}
\frac{\partial_u(\sqrt{-g}g^{uu}\partial_u\Psi^{(\pm)}_\omega)}{\sqrt{-g}g^{uu}}-\frac{g^{tt}}{g^{uu}}\left(\omega \pm A_0^{(3)}\right)^2 \Psi^{(\pm)}_\omega  + \frac{3}{R^2g^{uu}} \Psi^{(\pm)}_\omega = 0 \, ,
\end{equation}
which for the specific metric on the D7-brane world-volume become 
\begin{multline}
\label{e:scalarflucts}
\partial^2_u \Psi^{(\pm)}_\omega 
 + \frac{\partial_{u}\left(u^5\left(1+\frac{R^2}{4
     u^2}\right)\left(1-\frac{R^2}{4 u^2}\right)^3\right)}{u^5\left( 1
   + \frac{R^2}{4 u^2}\right)\left(1-\frac{R^2}{4
     u^2}\right)^3}\partial_{u} \Psi^{(\pm)}_\omega \\[1ex]
+ \left[\frac{R^4}{u^4\left(1+\frac{R^2}{4 u^2}\right)^2}\left(\omega
  \pm \mu \right)^2 
  + \frac{3}{u^2}\right] \Psi^{(\pm)}_\omega= 0 \, .
\end{multline}
These equations can be solved by reducing them to Schr\"{o}dinger
form. A very similar equation has been analysed in
\cite{Erdmenger:2010zm} for the determination of the mesonic spectrum
on the world-volume of a probe $D7$ brane at $T=0$, in global AdS
space.  Following steps similar to those in \cite{Erdmenger:2010zm},
and focusing on modes which are constant both on the $S^3\in S^5$ and
on the gauge theory $\bar{S}^3$, we obtain for the spectrum of
fluctuations
\begin{equation}
(\mu+\omega)R=\pm(3+2n) \, \quad n=0,1,2,\ldots\,.
\end{equation}
% This is Erdmenger/Filev (4.11) with l=lbar=0.
Here $n$ is the main quantum number. We see that the key effect of
the non-vanishing chemical potential is to shift the frequency $\omega
\rightarrow \omega + \mu$. Because of this we see that for large
enough chemical potential $\mu > \mu_{\text{crit}} = 3/R$, the frequency of
the lowest lying mode becomes zero, signalling that the homogeneous
solution potentially becomes unstable at this value of the chemical
potential, and a condensate of the scalar might form.

\subsection{Vector fluctuations at zero temperature}

Following the perturbative analysis in the scalar sector, we now turn
our attention to vectors.  We again expect unstable modes, but would
like to know whether or not they occur before the instability of the
scalar sector.  An analysis of the vector mode spectrum was performed
in infinite volume limit (on the Poincar\'e patch) in
\cite{Kruczenski:2003be} and then later extended to non-zero chemical
potential in \cite{Ammon:2008fc,Erdmenger:2008yj,Ammon:2009fe}. At
finite volume and vanishing temperature and chemical potential the
spectrum can be found in \cite{Erdmenger:2010zm}. The upshot of the
analysis of \cite{Kruczenski:2003be} is that the lowest lying
supermultiplet consists of two transverse scalars describing the
transverse fluctuations of the D7-brane in the $S^5$, one scalar which
originates from the vector component in the internal $S^3\in S^5$
wrapped by the D7-brane, and gauge components in the non-compact
directions of $\text{AdS}_5$.  As one moves to the compact case,
i.e.~global $\text{AdS}$ space~\cite{Erdmenger:2010zm}, the states
from this supermultiplet get reorganised (split) so that the lightest
state in the compact space is the scalar which originates from the
component of the gauge field in the direction of the internal $S^3 \in
S^5$. The vector components in the direction of the dual sphere as
well as the transverse scalars both have larger masses. We now want to
see how the fluctuations from the vector sector are shifted upon
introducing a chemical potential.

\subsubsection{The gauge theory vector fluctuations}

Let us start with the vector components in the direction of the sphere
$\bar{S}^3$ of the dual gauge theory.  These fluctuations are dual to
the vector excitations in the gauge theory.  Similarly to what we did
for scalars, we start by writing the fluctuations as
\begin{equation}
\label{vectorfluct}
A = A_0^{(3)}(u)\tau^3 {\rm d}t + R
a_i^{(1)}(t,u,\bar{\Omega}_3)\tau^1 {\rm d}\bar{\theta}^i + R
a_i^{(2)}(t,u,\bar{\Omega}_3)\tau^2{\rm d} \bar{\theta}^i \, ,
\end{equation}
where $i=(1,2,3)$ are indices on the dual $\bar{S}^3$, and as for
scalars we fluctuate in the gauge directions orthogonal to
$A_0^{(3)}$. Since we know that the lowest lying vector is a singlet on
$S^3$ we do not have to consider excitations which depend on the
coordinates of the internal sphere.\footnote{In the language of
  \cite{Erdmenger:2010zm} we consider type II fluctuations, with type
  I fluctuations to be considered in the next subsection. Looking
  ahead, it turns out that type III fluctuations condense after type~I
  fluctuations and we will not consider them in detail here.}

Next, we linearise the Yang-Mills action on the world-volume of the
D7-probe,
\begin{multline}
\label{eq:vectoractionYM}
S= -T_{D7}\pi^2\frac{R^4}{2\lambda} \int\!{\rm d}^8\xi\sqrt{-g}\bigg[(\partial_uA_0^{(3)})^2 g^{tt}g^{uu} \\[1ex]
+ R^2 \left(((D_ta_i)^{(a)})^2g^{tt}g^{ii}+(\partial_ua_i^{(a)})^2g^{uu}g^{ii}\right)
 + R^2 (f_{ij}^a)^2g^{ii}g^{jj}\bigg]\,,
\end{multline}
where
\begin{equation}
(D_ta_i)^{(a)} = \partial_ta_i^{(a)} - \epsilon^{abc}A_0^{(b)}a_i^{(c)} \,,\qquad
f_{ij}^a =  (\partial_ia_j^{(a)}-\partial_ja_i^{(a)})\, .
\end{equation}
Here, $g_{ij}$ is the metric on $\bar{S}^3$. The equations of motion
for the fluctuations $a_i^{(a)}$ are given by
\begin{multline}
	\sqrt{-g} \epsilon^{abc}A_0^{(c)}D_t a_i^{(b)}g^{tt}g^{ii}+\sqrt{-g}\partial_t\left(D_t a_i^{(a)}\right)g^{tt}g^{ii}+\partial_u\left(\sqrt{-g}\partial_u a_i^{(a)}g^{uu}g^{ii}\right)\\[1ex]+\sum_j\left(\sqrt{-g}\left(\partial_j a_i^{(a)}-\partial_i a_j^{(a)}\right)g^{jj}g^{ii}\right)=0,
\end{multline}
In order to solve this equation let us Fourier transform in the time direction.
%\begin{equation}
%a_i^{(a)}(t,u,\bar{\Omega}_3)= \int d \omega e^{-i\omega t}a_i^{(a)}(u,\omega,\%bar{\Omega}_3)
%\end{equation}
In order to decouple the equations for the fluctuations $a^{(1)}_i$ and
$a^{(2)}_i$ we introduce a new pair of variables
\begin{equation}
\label{linearsum}
\bar{X}_i^{(\pm)}(u,\omega,\bar{\Omega}_3)=e^{-i \omega t}\left( a_i^{(1)} (u,\omega,\bar{\Omega}_3) \pm ia_i^{(2)}(u,\omega,\bar{\Omega}_3) \right) \, .
\end{equation}
This finally yields the fluctuation equations
\begin{multline}
	\partial_u \left(\sqrt{-g}g^{uu}g^{ii}\partial_u
     \bar{X}_i^{(\pm)}\right)+\sum_j\partial_j\left(\sqrt{-g}g^{jj}g^{ii}\left(\partial_j
       \bar{X}_i^{(\pm)}-\partial_i
       \bar{X}_j^{(\pm)}\right)\right)\\[1ex]
-\sqrt{-g}g^{tt}g^{ii}\left(\omega \pm A_0^{(3)}\right)^2 \bar{X}_i^{(\pm)}=0\,,
\end{multline}
which are equivalent to 
\begin{equation}
  D^u \left(\partial_u \bar{X}_i^{(\pm)}\right)+ \nabla^j\left(\partial_j \bar{X}_i^{(\pm)}-\partial_i \bar{X}_j^{(\pm)}\right)- g^{tt} \left(\omega \pm A_0^{(3)}\right)^2 \bar{X}_i^{(\pm)}=0.
\end{equation}
In order to solve these equations, we make a factorised ansatz for
$\bar{X}_i$, as a product of radial and angular functions, and expand
the angular part $\bar{X}_i^{\bar{\Omega}_3}(\bar{\Omega}_3)$ on $\bar{S}^{3}$ in
terms of vector spherical harmonics. In general, the fluctuations
could also depend on a direction on internal sphere $S^3$. However, as
mentioned before, we will focus only on singlets under the global
$SO(4)$ symmetry group. Also, while there exists three type of vector
spherical harmonics on $\bar{S}^3$, it has been argued in
\cite{Kruczenski:2003be,Erdmenger:2010zm}, that for fluctuations of
the vector field which are taking place in $\bar{S}^3 \in
\text{AdS}_5$, only $Y_{i}^{l,\pm}$ which transform in
$\left(\left(\frac{l\mp 1}{2},\frac{l\pm 1}{2}\right) \right), l\geq
1)$ irreducible representations of $SO(4)$ are relevant.  Hence we
expand the fluctuations as
\begin{equation}
\label{expandvect}
\bar{X}_i^{(\pm)}(u,\omega,\bar{\Omega}_3) =\sum_{\bar{l},s=\pm} \bar{\Phi}_{\omega,\bar{l},s}^{(\pm)}(u) Y_i^{\bar{l},s} \, , 
\end{equation}
where the index $(\pm)$ refers to the two linear combinations of modes
as defined in (\ref{linearsum}), and the $\pm$ index refers to the
value of the index $s$ labelling the vector spherical harmonics.

The spherical harmonics satisfy the identities
\begin{equation}
\label{sphestuff}
\begin{aligned}
\nabla_i\nabla^i Y_j^{l,\pm}-R_j^k Y_k^{l,\pm} &= -(l+1)^2 Y_j^{l,\pm}\,, \\[1ex]
\epsilon^{ijk}\nabla_j Y_k^{l,\pm} &= \pm(l+1)Y_{l,\pm}^i\,, \\[1ex]
\nabla^i Y_i^{l,\pm} &= 0 \,,
\end{aligned}
\end{equation}
Using these identities the equation for the vector fluctuations can be
rewritten as
\begin{multline}\label{eq:vfeom}
\bar{\Phi}_{\omega,\bar{l},s}^{(\pm)''}(u)+\frac{\partial_u\left(\sqrt{-g(u)}g^{uu}(u) P(u)\right)}{\left(\sqrt{-g(u)}g^{uu}(u)P(u)\right)}\bar{\Phi}_{\omega,\bar{l},s}^{(\pm)'}(u)  \\[1ex]
-\frac{1}{g^{uu}(u)}\left[g^{tt}(u)\left(\omega \pm A_0^{(3)}(u)\right)^2+(\bar{l}+1)^2P(u)\right]\bar{\Phi}_{\omega,\bar{l},s}^{(\pm)}(u)=0 \, ,
\end{multline}
where the $\pm$ sign in front of $A_0$ in the equation is correlated
with the $(\pm)$ sign on the $\bar{\Phi}^{(\pm)}$ and $P(u)$ is the
inverse of the $u$-dependent part of the the metric factor in front of
${\rm d}\bar{\Omega}^2_3$.  We should note that this equation is
independent of the quantum number $s=\pm 1$, which will be different
when we start looking at the vector fluctuations in the direction of
the internal sphere. In the case of zero temperature the fluctuation
equation becomes (see also~\cite{Erdmenger:2010zm})
\begin{multline}
\label{eomvectorYMT0}
\bar{\Phi}_{\omega,\bar{l}}^{(\pm)''}(u)+\frac{\partial_u\left(u^3\left(1+\frac{R^2}{4u^2}\right)\left(1-\frac{R^2}{4u^2}\right)\right)}{u^3\left(1+\frac{R^2}{4u^2}\right)\left(1-\frac{R^2}{4u^2}\right)}\bar{\Phi}_{\omega,\bar{l}}^{(\pm)'}(u)\\[1ex]
+\left[\frac{R^4}{u^4\left(1+\frac{R^2}{4u^2}\right)^2}\left(\omega \pm \mu\right)^2-\frac{R^2}{u^4\left(1-\frac{R^2}{4u^2}\right)^2}(\bar{l}+1)^2\right]\bar{\Phi}_{\omega,\bar{l}}^{(\pm)}(u)=0 \, .
\end{multline}
This equation is very similar to \eqref{e:scalarflucts}, and can again
be cast in Schr\"odinger form. We then find that the spectrum of
vector fluctuations for the $\bar{l}$-th spherical harmonics is given
by
\begin{equation}
\label{spectrum}
(\omega \pm \mu)R = (3+\bar{l}+2n) \, \quad n= 0,1,2,3....\, \quad \bar{l}=1,2,3.... \, , 
\end{equation}
% This is Erdmenger/Filev (4.37 with l=0 (which is their ltilde=0)
% type II modes
Here $n$ is again the main quantum number, and $\bar{l}$ is an $SO(4)$
quantum number corresponding to the sphere $\bar{S}^3\in
\text{AdS}_5$. We see that the result is again the same as the one in
\cite{Erdmenger:2010zm} if we consider $SO(4)$ singlets (i.e.~set
$l=0$), except that the chemical potential shifts the frequency
$\omega \rightarrow \omega \pm \mu$.

From equation \eqref{spectrum} we see that for a critical value of
the chemical potential given by $\mu_{\text{crit}}=4/R$, the lowest lying mode
$\bar{l}=1$ will become massless. Therefore, we expect that when the
chemical potential is larger than this value, the system potentially
becomes unstable.

\subsubsection{The charged scalar fluctuations}
\label{s:T0dualscalar}

Let us now consider fluctuations of the vector field in the direction
of the internal $S^3 \in S^5$, which are dual to an $SO(4)$ charged
scalar field in the gauge theory. Since the WZW term in the action is
now non-zero, when considering fluctuations, we have to modify the
action from (\ref{eq:vectoractionYM}) by adding the term
\begin{equation}
S_{WZW} =\frac{T_{D7}\pi^2R^4}{\lambda}\int \text{Tr}(C\wedge F\wedge F)\,,\quad
{\rm d}C =\frac{4}{R}\text{Vol}(\text{AdS}_5).
\end{equation}
Similarly as before, we make an ansatz as in (\ref{vectorfluct})
except that the index $i$ is now taking values in the internal $S^3$. In
addition, we will also allow the fluctuations to depend on both $S^3$ and
$\bar{S}^3$ variables.  In order to decouple the equations of motion
we redefine variables as in (\ref{linearsum}) and make a factorised
ansatz
\begin{equation}
X_{i}^{(\pm)}(u,\omega,\Omega_3,\bar{\Omega}_3) = e^{-i\omega t} \Phi^{(\pm)}(u)\bar{Y}^{\bar{l}}(\bar{\Omega}_3) Y^{l,s}_i(\Omega_3) \, ,
\end{equation}
where the index $(\pm)$ refers to the sign in the linear combination
(\ref{linearsum}), $i$ denotes the index in the direction of the
internal $S^3$, and the index $s=\pm 1$. Also to shorten the notation
we have suppressed indices on the functions $\Phi^{(\pm)}$, which
should really also carry indices $(\omega,s,\bar{l},l)$.

Following the same procedure as for scalars and the other gauge
components we arrive at the equation for the fluctuations (see also~\cite{Erdmenger:2010zm})
\begin{multline}
\label{vectscal}
\partial_u^2  \Phi^{(\pm)} + \frac{\partial_u(u^5\left(1+\frac{R^2}{4u^2}\right)\left(1-\frac{R^2}{4u^2}\right)^3)}{u^5\left(1+\frac{R^2}{4u^2}\right)\left(1-\frac{R^2}{4u^2}\right)^3}\partial_u \Phi^{(\pm)} + \frac{R^2}{u^4\left(1+\frac{R^2}{4u^2}\right)^2}(\omega\pm \mu)^2 \Phi^{(\pm)}  \\[1ex]
-\left[(l+1)^2+ \bar{l}(\bar{l}+2)\frac{R^2}{u^2\left(1-\frac{R^2}{4u^2}\right)^2}+4s (l+1)\right]\frac{1}{u^2} \Phi^{(\pm)} = 0 \, , 
\end{multline}
where the sign in the $(\omega \pm \mu)$ is the same as for
$\Phi^{(\pm)}$. We should note that this equation explicitly depends
on the quantum number $s$, which is labelling the vector harmonics on
$S^3$. This is in contrast to the previous case for the equation for
vector fluctuations on $\bar{S}^3$.

Putting equation (\ref{vectscal}) in Schr\"{o}dinger form,
like we did for the other fluctuations, we obtain for the spectrum
\begin{multline}
  (\omega\pm\mu)R=3+2s+l+2n+\bar{l} \,  \quad \quad  \text{where}  \\[1ex]
  \bar{l} = 0,1,2,\ldots \quad  l = 1,2,3,\ldots\quad  n=0,1,2,\ldots\quad  s=\pm 1 \, .
\end{multline}
% This is Erdmenger/Filev (4.31), type I modes
This is the same as (4.31) of \cite{Erdmenger:2010zm} except for the
shift $\omega \rightarrow \omega +\mu$. Let us also note that the
lowest lying excitation carries quantum numbers $(\bar{l}=0,l=1, s=-1,
n=0)$, and this mode will reach zero frequency when
$\mu>\mu_{\text{crit}}=2/R$.\footnote{The fact that this excitation has the
lowest mass at $\mu=0$ was also observed in~\cite{Erdmenger:2010zm}, but no
attempt was made to study its condensation under the influence of a
chemical potential.}

\bigskip

\section{Perturbative analysis of the homogeneous vacuum at $T\not=0$}
\label{perturbfinite}

In the previous section we have observed that the homogeneous
isotropic ground state at non-zero chemical potential and zero
temperature was unstable under both scalar and vector fluctuations. We
would now like to see how is this modified once the temperature is
turned on, paying particular attention to the order in which the
instabilities set in as a function of temperature. 

We should emphasise that in contrast to the zero temperature case,
where all fluctuations have real frequency $\omega$ (corresponding to
stable mesonic scalar and vector particles), at finite temperature
(above the Hawking-Page transition), even in the absence of chemical
potential all fluctuation frequencies have a non-vanishing imaginary
part. When the chemical potential is zero the imaginary part of these
frequencies are negative, corresponding to the fact that these
excitations are decaying in time, i.e.~that they describe
\emph{quasi-stable} particles. However, as the chemical potential is
turned on, if there is indeed an instability present, we expect that
the negative imaginary part of the frequencies will become positive,
i.e.~that a decaying excitation would become an exponentially growing
mode, which signals an instability. In what follows, we will therefore
focus on studying the imaginary part of the quasi-normal modes of the
system.

We start our analysis by looking at the scalar fluctuations, and
repeat the procedure similar to that at zero temperature. Again, we
use the metric on the $S^5$ as in (\ref{S5metrics}), keeping only the
transverse scalar $\Psi(t,u)$ nonzero (see
equation~\eqref{e:Psitu_def}) and making it depend only on time and
the radial direction $u$. As argued before, such excitation is
consistent with the full equations of motion, and should correspond to
the lowest energy mode, an S-wave on $\bar{S}^3$. The induced metric
on the D7-brane world-volume is then
\begin{multline}
{\rm d}s^2=-\frac{2\rho_H^2 F(u)}{u R^2 W(u)}{\rm d}t^2 +
2\frac{\rho_H^2}{u}W(u){\rm d}\bar{\Omega}_3^2 
+\frac{R^2}{4u^2 F(u)}{\rm d}u^2\\[1ex] 
+ R^2\left(1-\frac{\Psi(t,u)^2}{R^2}\right){\rm d}\Omega^2_3
+\left(\frac{\partial \Psi(t,u)}{\partial t}{\rm d}t+\frac{\partial
    \Psi(t,u)}{\partial u}{\rm d}u\right)^2 \, ,
\end{multline}
where
\begin{equation}
F(u)= 1 - u^2 \, , \quad W(u) = 1 - \frac{u R^2}{4 \rho_H^2} \, .
\end{equation}
As before, we Fourier transform the scalar $\Psi(t,u)$ and make the
ansatz that it is pointing in the direction orthogonal to the $A_0^{(3)}
\tau^3$ in colour space,
\begin{equation}
\Psi(t,u) = \int\! \frac{{\rm d}\omega}{2 \pi} e^{-i\omega t} (\Psi^{(1)}_\omega(u)\tau^1+\Psi^{(2)}_\omega(u)\tau^2) \, .
\end{equation}
We then again change variables as
\begin{equation}
\Psi_\omega^{(\pm)}(u) =  \Psi^{(1)}_\omega(u) \pm i \Psi^{(2)}_\omega(u) \, ,
\end{equation}
so that the equations for the $\Psi^{(\pm)}$ fluctuations decouple and
are given by\footnote{Note that at $T=0$, the equations only depend on
  $\omega-\mu$, and hence the critical chemical potential coincides
  with the frequency of the lightest mode. At $T>0$ the $A^{(3)}_0$
  component is no longer a constant, and obtaining the critical
  chemical potential is more complicated (physical states are no
  longer straight lines in the $\omega,\mu$ plane).}
\begin{multline}
\label{fluctuationscalbh}
\frac{\partial_u \left(\frac{W(u) F(u)}{u}\partial_u
    \Psi_{\omega}^{(\pm)}(u)\right)}{\frac{W(u) F(u)}{u}}+\frac{ R^4
  W(u)}{8u\rho_H^2 F(u)^2}\left(\omega  \pm A_0^{(3)}(u)\right)^2
\Psi_{\omega}^{(\pm)}(u)\\[1ex]
+\frac{3}{4u^2 F(u)}\Psi_{\omega}^{(\pm)}(u) = 0\, .
\end{multline}
We solve the fluctuation equation by imposing that the modes satisfy
an incoming boundary condition at horizon,
\begin{equation}
\Psi_\omega^{(\pm)}(u)\Big|_{u\approx 1}= (1-u)^{-i\omega/(4\pi T)}(1+O(1-u)) \, .
\end{equation}

\begin{figure}[t]
\begin{center}
  \includegraphics[width=.48\textwidth]{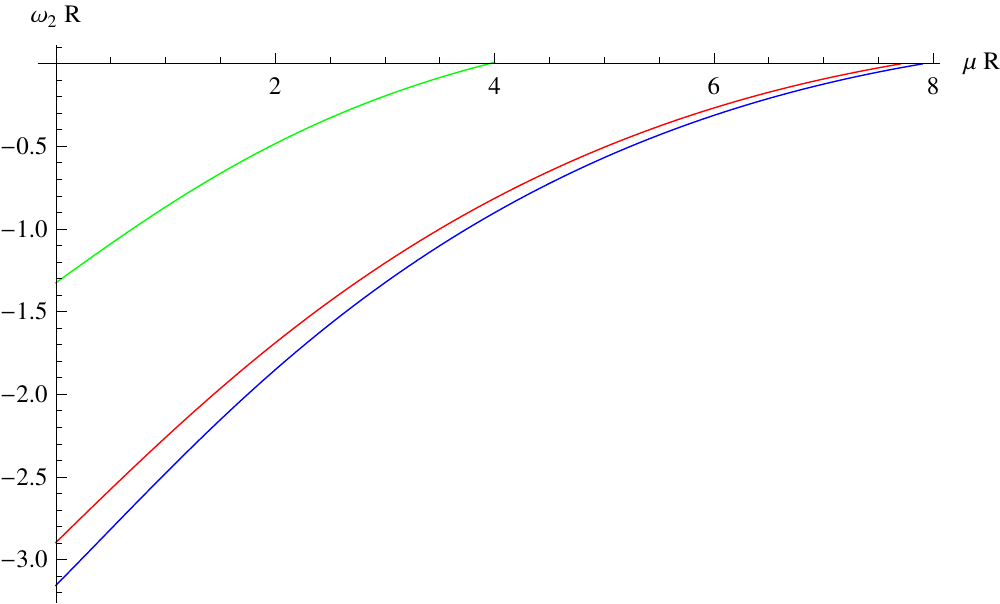}\quad
\includegraphics[width=.48\textwidth]{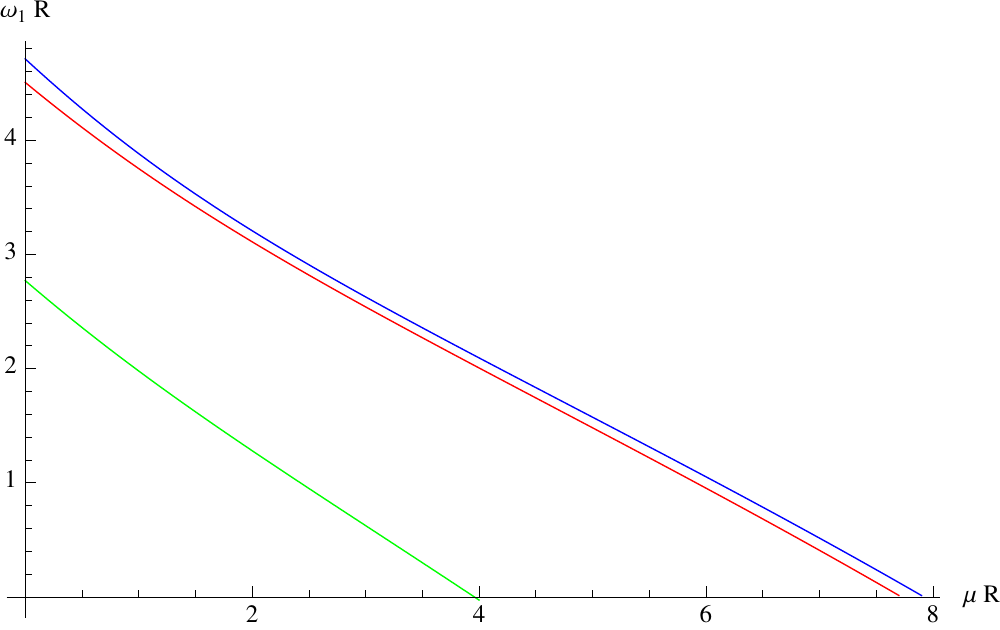} 
\end{center}
\caption{Plots of the imaginary (left) and real (right) parts of the
  frequencies for the lowest lying uncharged scalar fluctuation (red), the
  vector fluctuation (blue) and the charged scalar fluctuation (green), at fixed temperature $\pi TR=2$,
  as a function of the chemical potential.
\label{frequenciesmu}
}
\end{figure}

The equations for the vector fluctuations are derived in a similar
way.  Let us first consider the vector fluctuations which are dual to
vectors. We take them to be singlets under the global $SO(4)$ ($l=0$),
and orthogonal to the isospin chemical potential in the gauge group as
we did at zero temperature, see (\ref{vectorfluct}).  Following steps
similar to those at zero temperature, instead of
$a_i^{(1)},a_i^{(2)}$, we introduce a new pair of variables
$X_i^{(\pm)}$, as in (\ref{linearsum}) and Fourier expand it in
spherical harmonics as in (\ref{expandvect}). Hence, we arrive at the
equations of motion for these fluctuations
\begin{multline}
\label{fluctvectbh}
\partial^2_u \bar{\Phi}_{\omega}^{(\pm)s,\bar{l}}
+ \frac{\partial_u F(u) \partial_u \bar{\Phi}_{\omega}^{(\pm)s,\bar{l}}}{F(u)}
-(\bar{l}+1)^2\frac{R^2}{8\rho_H^2 W(u) F(u)u}\bar{\Phi}_{\omega}^{(\pm)s,\bar{l}}\\[1ex]
+\frac{R^4W(u)}{8u\rho_H^2F(u)^2}(\omega\pm A_0^{(3)})^2 \bar{\Phi}_{\omega}^{(\pm)s,\bar{l}} = 0\, .
\end{multline}
As for scalars, we impose incoming boundary conditions at the black hole horizon
\begin{equation}
\bar{\Phi}_{\omega}^{(\pm)s,\bar{l}}  \Big|_{u\approx 1}= (1-u)^{-i\omega/(4\pi T)}(1+O(1-u)) \,.
\end{equation}
Finally, we turn to the vector fluctuations dual to the charged
scalars. Following similar steps as we did at zero temperature we
arrive at the equations governing these fluctuations (see equation
(\ref{vectscal}))
\begin{multline}
\label{fluctvectscalbh}
\frac{\partial_u\bigg(\frac{W(u)F(u)}{u}\partial_u\Phi^{(\pm)}\bigg)}{\frac{W(u)F(u)}{u}}
-(l+1)^2\frac{1}{4u^2F(u)} \Phi^{(\pm)} 
-\bar{l}(\bar{l}+2)\frac{R^2}{8\rho_H^2W(u)F(u)u} \Phi^{(\pm)}  \\[1ex]
-\frac{s(l+1)}{u^2F(u)} \Phi^{(\pm)} 
+\frac{R^4W(u)}{8u\rho_H^2 F(u)^2}(\omega\pm A_0^{(3)})^2 \Phi^{(\pm)} = 0 \, .
\end{multline}
where we have again suppressed the indices $(\omega,\bar{l},l,s)$ on
the functions $\Phi$ and we impose incoming boundary conditions at the
horizon of the black hole
\begin{equation}
\Phi^{(\pm)}\Big|_{u\approx 1}=(1-u)^{-i\omega/(4\pi T)}(1+O(1-u))\,.
\end{equation}
In order to solve the fluctuation equations \eqref{fluctuationscalbh},
\eqref{fluctvectbh} and \eqref{fluctvectscalbh}, we use a shooting
technique, in which we start from the horizon and look for modes that
decay at infinity i.e.~we look for the modes that describe
normalisable excitations. These boundary conditions will be satisfied
only for a discrete set of frequencies. We plot the imaginary parts of
those frequencies for the scalar and two vectors, for fixed temperature
and various values of the chemical potential $\mu$, in figure~\ref{frequenciesmu}.

We see that as the value of the chemical potential is increased, the
imaginary parts of the frequencies, which were initially all negative,
become less and less negative and approach zero. When the chemical
potential exceeds a critical value, the imaginary parts become
positive one by one, signalling the presence of unstable modes in the
system. Similarly, the real parts of the frequencies are decreasing to
zero as the chemical potential grows, signalling again the onset of an
instability. For a particular value of the temperature presented on
the left plot in the figure \ref{frequenciesmu}, we see that the
vector dual to the gauge theory charged scalar remains the lightest in the
spectrum and condenses first, followed by the transverse scalar and
finally the vector.
\begin{figure}[t]
\begin{center}
\includegraphics[width=.46\textwidth]{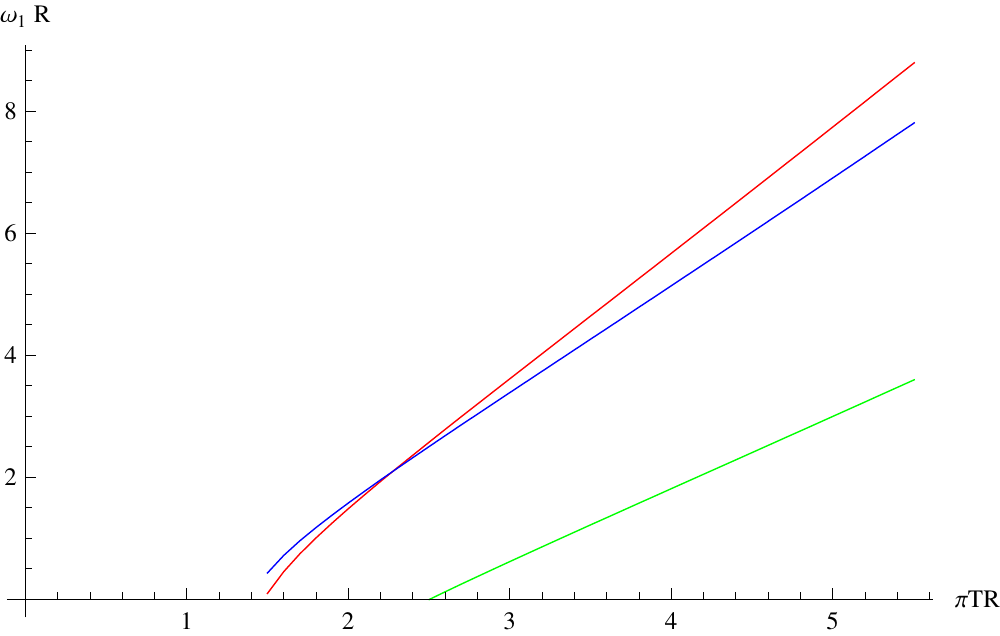}\quad
\includegraphics[width=.5\textwidth]{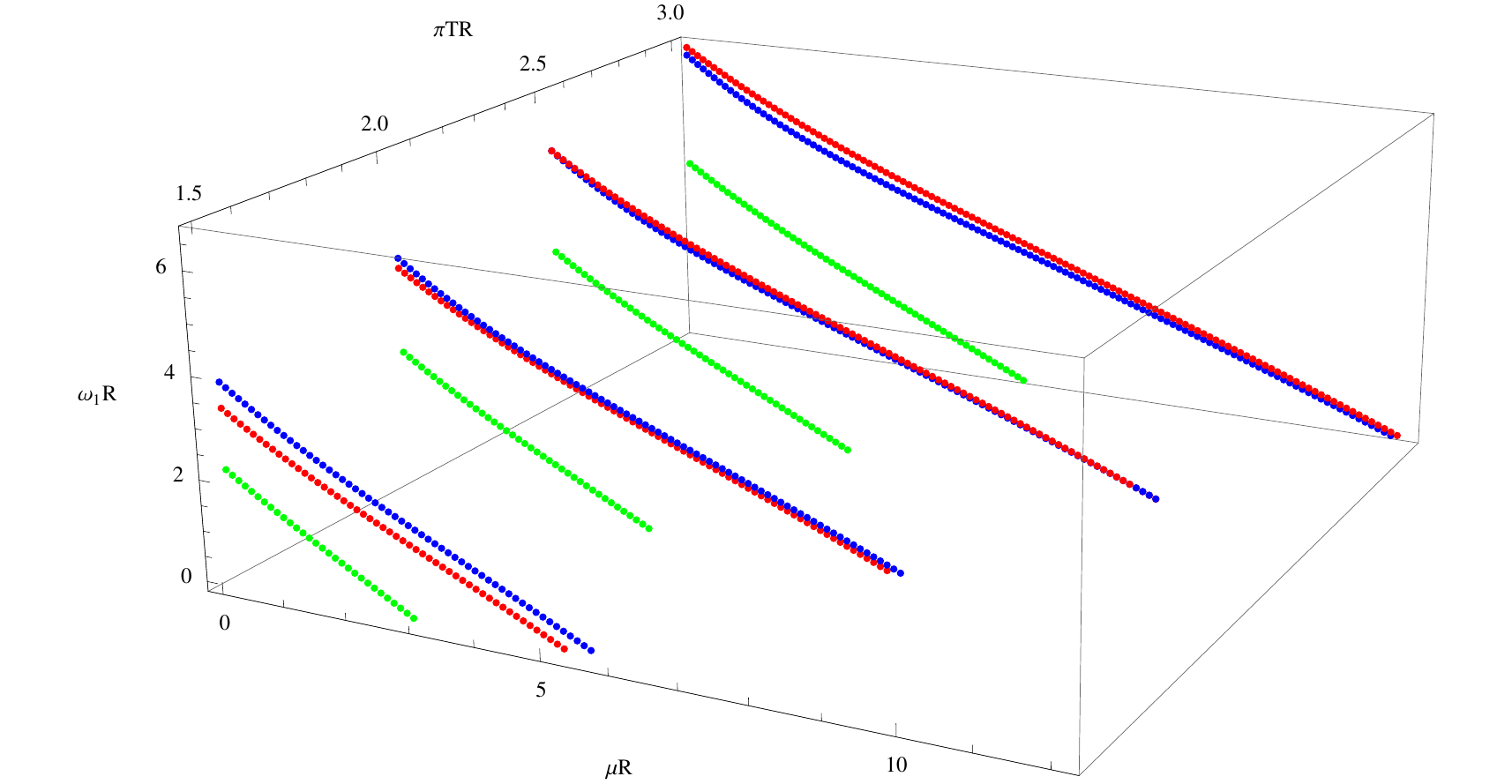}
\end{center}
\caption{Left is a plot of the real parts of the frequencies for the
  various modes (colours as in figure~\ref{frequenciesmu}) as
  functions of the temperature at fixed value $\mu R = 5$ of the
  chemical potential.  The plot on the right shows the real parts of
  the frequencies as functions of both temperature and chemical
  potential.\label{frequenciesmusec}}
\end{figure}

\begin{figure}[t]
\begin{center}
\includegraphics[width=.49\textwidth]{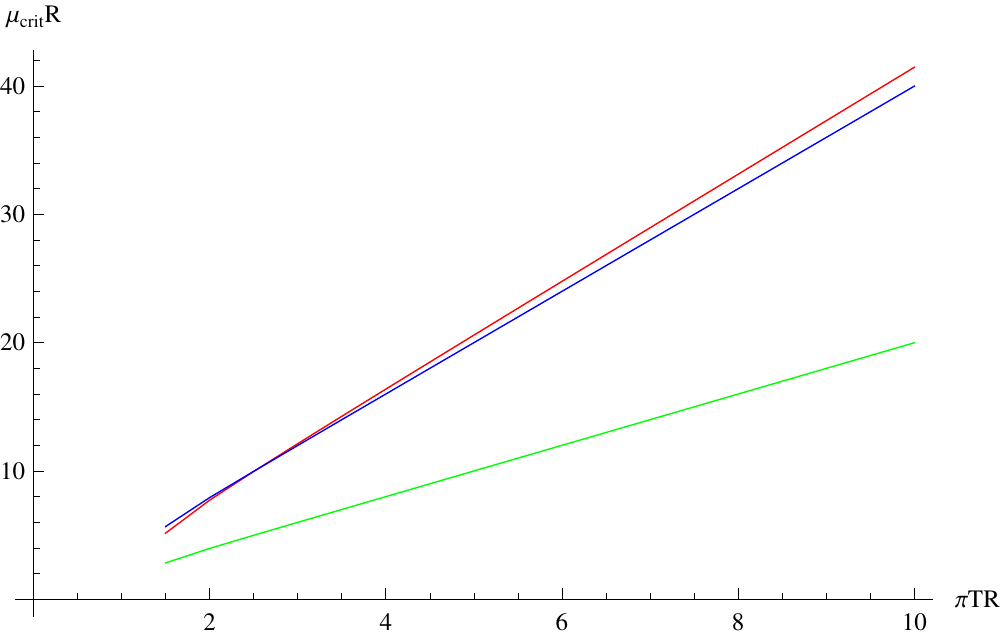}\quad
\includegraphics[width=.47\textwidth]{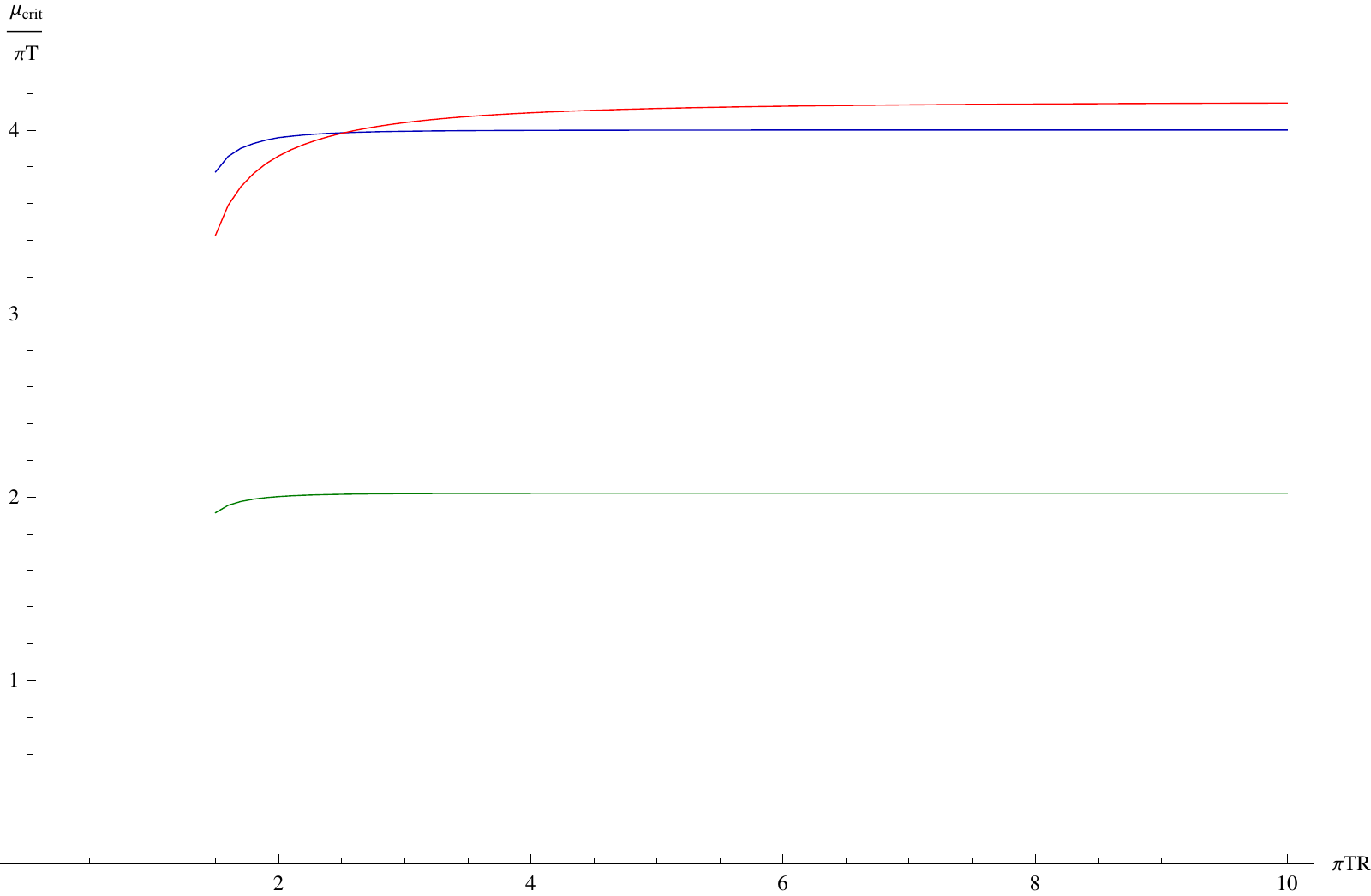}
\end{center}
\caption{Critical chemical potential as function of temperature, in
  two different dimensionless combinations. The figure on the right
  shows more clearly what happens in the $TR\rightarrow\infty$ limit, which can be
  interpreted as the large radius limit at fixed temperature.
  \label{valueofmus}}
\end{figure}

In general, one would expect that particles condense roughly when the
chemical potential become of the order of their mass. It is thus of
interest to look at the behaviour of the masses\footnote{We use pole
  masses here for convenience as they are easy to obtain from the
  quasi-normal mode analysis, and the intuition we want to verify is
  anyhow qualitative.} as a function of
temperature. Figure~\ref{frequenciesmusec} shows the result of this
analysis. We here observe another interesting phenomenon, namely that
there is a crossover point at some critical value of the temperature,
above which the lightest vector becomes lighter than the transverse
scalar. This suggests that above the crossover temperature, the
lightest vector would condense before the lightest transverse scalar,
if it had not been for the $SO(4)$ charged scalar that condenses even
earlier. One can indeed see that the corresponding imaginary parts
cross as well, approximately at this point, see figure~\ref{valueofmus}.

%to the vector, remains lightest as the temperature is varied. This
%suggests that for all temperatures, a new ground state is formed by
%condensation of the vector component dual to a gauge theory scalar.
%We will check this fact in the full non-linear theory in the following
%section.

For large $TR$, the results read
\begin{equation}
\begin{aligned}
\text{charged scalar}:&&\qquad \mu_{\text{crit}}R & \approx 2.00\pi TR -
\frac{2.00\times 0.05}{4\pi TR} + \cdots \,,\\[1ex]
\text{vector}:&&\qquad \mu_{\text{crit}}R&\approx 4.00\pi TR-\frac{4.00 \times 0.05}{4\pi TR}+\cdots\,,\\[1ex]
\text{uncharged scalar}:&&\qquad \mu_{\text{crit}}R&\approx 4.16\pi TR-\frac{4.16\times 1.00}{4\pi TR}+\cdots\,.
\end{aligned}
\end{equation}
The leading order terms should agree with those obtained in the
Poincar\'e patch, though to our knowledge only the one for the vector
has been computed in the
literature~\cite{Erdmenger:2008yj,Ammon:2009fe,Peeters:2009sr}. The
result for the critical chemical potential
of~\cite{Erdmenger:2008yj,Ammon:2009fe} (when extrapolated to zero bare
quark mass) seems to be somewhat larger than ours, which may be due to
the fact that we have used a Yang-Mills truncation rather than the
full DBI action.

\section{The new ground states at zero and finite temperature}

In the previous two sections we have seen that for large enough
chemical potential the homogeneous ground state on $\bar{S}^3 \in
\text{AdS}_5$ becomes unstable under both scalar and various components of
vector fluctuations. This is happening both at zero and non-zero
temperature.  In particular we observe that, at zero temperature,
vector fluctuations in the direction of the internal $S^3 \in S^5$ are
the first to became unstable. As the temperature is increased, these
vector components remain the first to become unstable. On the other
hand, the ordering in which the other components of the vector
fluctuations and the scalar fluctuations become unstable is dependent
on the temperature, as there is a `crossover' temperature above which
all vector components first become unstable.

%Hence, the main difference that turing on temperature introduces is
%that at zero temperature the scalars are unstable at lower value of
%chemical potential than vectors, while at finite temperature there is
%crossover point, above which the vectors are first to become unstable.

Our previous analysis was done in perturbation theory, i.e.~at the
linearised level.  So we would now like to see if the instabilities
which we have found are present in the full non-linear theory, and to
explore the new ground state in which the system settles for large
enough values of the chemical potential.

\subsection{The new ground state at zero temperature}

In section~\ref{s:T0dualscalar} we have observed that as the chemical
potential is turned on, when its value reaches $\mu\geq 2/R$, the
lowest lying mode of the vector component in the internal $S^3$,
becomes massless, signalling the onset of possible instability in the
system.  We have also seen that for even larger values of the chemical
potential, the scalar becomes massless at $\mu \geq 3/R$ and the other
components of the vector develop an instability for $\mu\geq 4/R$.

To see whether the appearance of these massless modes indeed signals a
real instability, we will now turn to the full non-linear theory and
try to explicitly construct the new ground state to which the system
would evolve as a consequence of the instability.  As the perturbative
analysis suggests that vector components in the internal $S^3$
direction are first to condense, we will start the analysis of the new
ground state by turning on only those components. We will later, for
comparison, also analyse possible ground states due to condensation of
the other fluctuations, and verify that those always have higher
energy than the scalar condensate.

When writing down an ansatz for the scalar condensate ground state, we
will use the fact that in perturbation theory, the first unstable mode
is an $\bar{l}=0,l=1, n=0,s=-1$ wave (where $\bar{l}$ labels modes in
the $\bar{S}^3$ and $l$ labels modes in the $S^3 \in S^5$). As far as
the $A_0$ component is concerned, at linearised level one cannot see
the back-reaction of the scalar on the background value of this field,
so in principle one cannot say if in the new ground state the $A_0$
component will start to depend on the angular coordinates or not.

As a simplest attempt we take $A_0$ to remain homogeneous,
i.e.~independent of the $\bar{S}^{3}$ angular coordinates. With this
ansatz, potential problems in the equations of motion could originate
from expression of the form ``$A_{\alpha}A_{\beta}g_{S^3}^{\alpha
  \beta}$'', which are now turned on due to the non-vanishing vector
field in the direction of the internal sphere $S^3$.  Since these
terms will typically produce spherical harmonics of higher $l$-number
we would need to balance them in the equations of motion. However, we
expect that the ground state would originate from condensation of only
the lowest harmonic, so that higher $l$-harmonics are not needed. It
is possible to reconcile these two observations if the
``$A_{\alpha}A_{\beta}g_{S^3}^{\alpha\beta}$'' expression is
independent of the angular coordinates. This can indeed be achieved
for a particular linear combination of spherical harmonics given by
\begin{multline}
\label{harmonicslinear}
Y_{\alpha} = \frac{i k_0}{K} Y_{\alpha}^{1,0,0,-1} + \frac{(k_1 + i k_2)}{K} Y_{\alpha}^{1,0,-1,-1}   + \frac{(k_1 - i k_2)}{K} Y_{\alpha}^{1,0,1,-1}   \\[1ex]
\text{where}\qquad K \equiv \sqrt{k_0^2 + 2(k_1^2 + k_2^2)}\,,
\end{multline}
and $k_0,k_1,k_2$ are three arbitrary real numbers which are not
simultaneously vanishing, and $(l,m_1,m_2,s)$ are the quantum numbers
of the spherical harmonics. We should note here that value of these
quantum numbers will be taken to be the same as those of the
lowest lying excitation we have previously found in the perturbative
analysis.  Explicitly, the spherical harmonics are given by
\begin{equation} 
\begin{aligned}
Y^{1,0,0,-1} &= \frac{i}{2}{\rm d}{\alpha} + \frac{i}{2}\cos \beta {\rm d}\gamma   \\[1ex]
Y^{1,0,1,-1} &= -\frac{1}{2\sqrt{2}}e^{-i \alpha}{\rm d}\beta-\frac{i}{2\sqrt{2}}\sin\beta e^{-i \alpha} {\rm d}\gamma  \\[1ex]
Y^{1,0,-1,-1} &= -\frac{1}{2\sqrt{2}}e^{i\alpha}{\rm d}\beta + \frac{i}{2\sqrt{2}}\sin\beta e^{i \alpha} {\rm d}\gamma \, ,
\end{aligned}
\end{equation}
where $\alpha,\beta,\gamma$ are Euler coordinates on $S^3 \in S^5$,
\begin{equation}
{\rm d}s^2_{S^3}=\frac{1}{4}({\rm d}\alpha^2+{\rm d}\beta^2+{\rm
  d}\gamma^2+2\cos\beta {\rm d}\alpha {\rm d}\gamma) \,. 
\end{equation}
It is also useful to keep in mind that
\begin{equation}
(Y^{l,m_1,m_2,s}_i)^*=-(-1)^{m_1-m_2}Y^{l,-m_1,-m_2,s}_i \, .
\end{equation}

Our ansatz for the new ground state is
\begin{equation}
\label{fullsolution}
A_0 = A_0^{(3)}(u) \tau^3\,,\qquad
A_{\alpha} = R \eta(u) Y_{\alpha}(\Omega_3) \tau^1
\end{equation}
Plugging this into equations of motion, and using identities
(\ref{sphestuff}) we get an equation for $A_0(u)$
\begin{equation}
\label{eomfullnonlinear1}
\frac{\partial_u\left((\partial_uA_0^{(3)})u^3\left(1-\frac{R^2}{4u^2}\right)^3/\left(1+\frac{R^2}{4u^2}\right)\right)}{u^3\left(1-\frac{R^2}{4u^2}\right)^3/\left(1+\frac{R^2}{4u^2}\right)}-\frac{R^2}{u^2}(\eta(u))^2A_0^{(3)}(u) = 0 \, , 
\end{equation}
and an equation for the function $\eta(u)$
\begin{equation}
\label{eomfullnonlinear2}
\frac{\partial_u\left(\partial_u\eta(u)u^5\left(1+\frac{R^2}{4u^2}\right)\left(1-\frac{R^2}{4u^2}\right)^3\right)}{u^5\left(1+\frac{R^2}{4u^2}\right)\left(1-\frac{R^2}{4u^2}\right)^3}
+\frac{R^4}{u^4\left(1+\frac{R^2}{4u^2}\right)^2}(A_0^{(3)}(u))^2\eta(u)+\frac{4}{u^2}\eta(u) = 0 \, .
\end{equation}
Note that these are independent of the parameter $K$, which only
appears in the angular part of the equations of motion, which is
automatically satisfied for our ansatz.

\begin{figure}[t]
\begin{center}
\includegraphics[width=.48\textwidth]{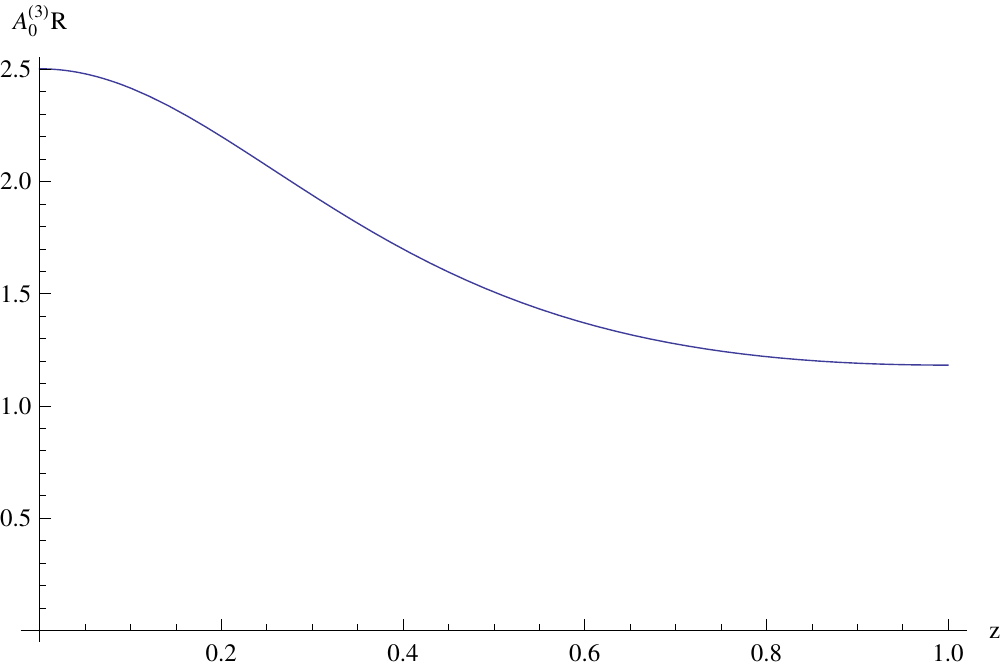} \quad
\includegraphics[width=.48\textwidth]{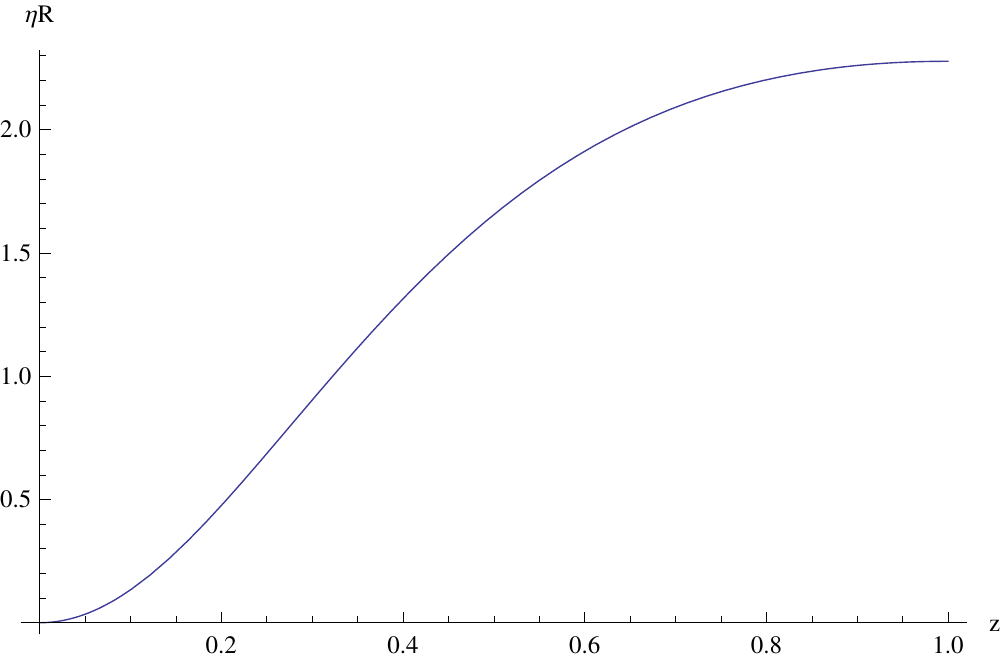}
\end{center}
\caption{Profile of the fields $A_0$ (left) and $\eta$ (right) of the
  charged scalar condensate, evaluated at $\mu R=2.5$. The boundary is at
  $z=0$ and the AdS centre is at $z=1$.\label{figuresolution}}
\end{figure}

Equations (\ref{eomfullnonlinear1}), and (\ref{eomfullnonlinear2}) are
written in the non-compact coordinate $u$ for which the AdS centre is
at $u=R/2$ and the boundary is at $u=\infty$.  However, for numerical
considerations it is more convenient to perform a coordinate change to
compact coordinates $z=R/2u$, so that the AdS origin is at $z=1$, and
the boundary at $z=0$. The equations of motion then are given by
\begin{equation}
\begin{aligned}
\label{fulleqnvec}
\partial_z^2A_0^{(3)}(z)+\frac{1+8z^2+3z^4}{z^5-z}\partial_z A_0^{(3)}(z)-\frac{R^2}{z^2} \eta(z)^2A_0^{(3)}(z) &= 0 \, , \\[1ex]
\partial_z^2\eta(z)+\frac{3+4z^2+5z^4}{z^5-z}\partial_z\eta(z)+\left(\frac{4}{z^2}+\frac{4R^2}{(1+z^2)^2}A_0^{(3)}(z)^2\right)\eta(z) &= 0 \, .
\end{aligned}
\end{equation}
We are interested in the solutions of these equations that are
regular everywhere, and in particular at the origin of AdS
space. This removes half of the solutions, as can be seen by looking
at the $z\rightarrow 1$ limit of the above equations. Namely, assuming
that $A_0$ and $\eta$ are regular at the AdS origin, it is easy to see
that the above equations reduce to the conditions that the first
derivatives of $A_0$ and $\eta$ are vanishing at the origin. Hence,
the general regular solution will be parametrised by two parameters
$a$, $b$. We then solve the equations of motion by shooting from the
AdS origin, and look for the solutions at the boundary such that
$\eta$ is normalisable, while $A_0$ is not. This normalisability
condition further reduces the number of parameters by one. Hence in the
expansion near infinity
\begin{equation}
\label{e:densities_definition}
A_0^{(3)} = \mu - \rho z^2 + \cdots \, , \quad A_{\alpha}^{(1)} = R \rho_\eta Y_{\alpha} z^2 + \cdots \, .
\end{equation}
both the densities $\rho$ and $\rho_\eta$ are functions of
the chemical potential $\mu$.

\begin{figure}[t]
\begin{center}
\includegraphics[width=.48\textwidth]{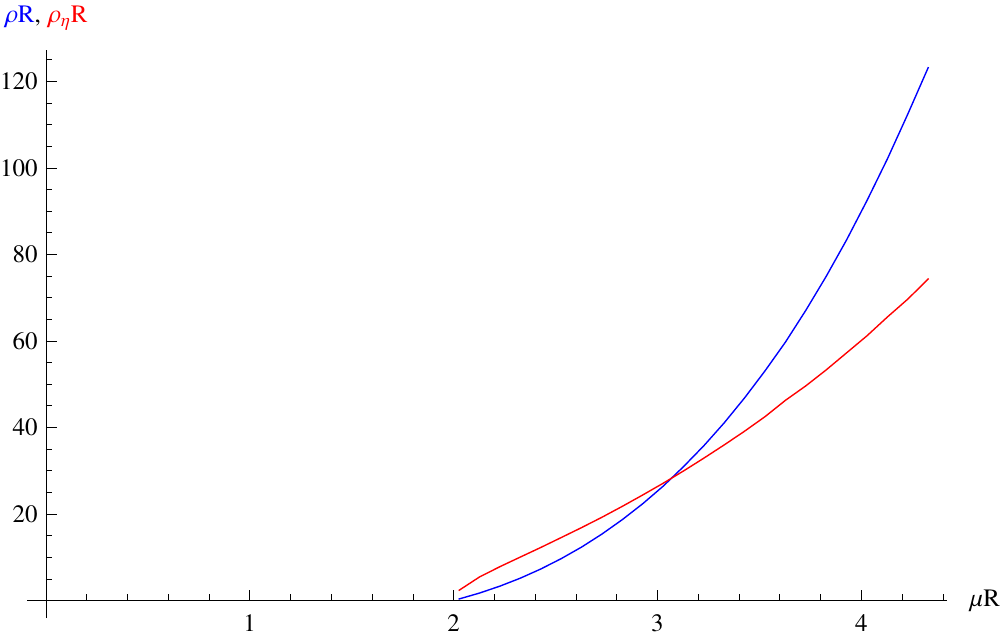} \quad
\includegraphics[width=.48\textwidth]{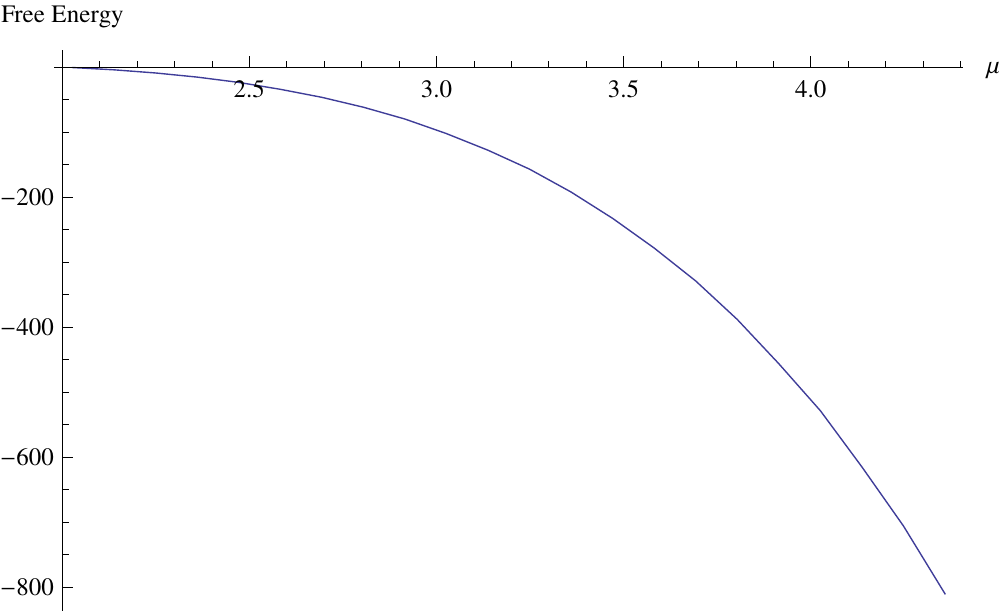}
\end{center}
\caption{The plot on the left shows the isospin density $\rho R$
  (blue) and scalar density $\rho_\eta R$ (red) of the charged scalar
  condensate as functions of the isospin chemical potential.  The plot
  on the right shows the scaled free energy as function of chemical
  potential.\label{solutionrhofnmus}}
\end{figure}

We plot the radial profile of the functions $A_0(z)$ and $\eta(z)$ for one
particular solution in figure~\ref{figuresolution}.  As required, we
see that the solution is regular everywhere, and approaches the origin of AdS
with vanishing derivative, so that no cusp is present.
We also study various solutions for different values of chemical
potential, see figure \ref{solutionrhofnmus}. The shooting procedure
shows that there is a critical value of the chemical potential
$\mu_{\text{crit}} \sim 2/R$ below which there is no nontrivial solution
present. Above $\mu=\mu_{\text{crit}}$ a nontrivial condensate of scalar
particles forms, and in the neighbourhood of $\mu_{\text{crit}}$, this
condensate is to a good approximation given by
\begin{equation}
\rho_\eta = \begin{cases} 0 & \text{for $\mu < \mu_{\text{crit}}$}\\
\sqrt{\mu - \mu_{\text{crit}}} & \text{for $\mu > \mu_{\text{crit}}$} \, .
\end{cases}
\end{equation}

We have also evaluated the free energy for various values of the
chemical potential (see figure~\ref{solutionrhofnmus}), and observed
that it is less than the (vanishing) free energy of the trivial
configuration, which is in agreement with the statement that this is
the ground state.

In summary, our analysis shows that for large enough value of the
chemical potential, this system undergoes a second order phase
transition in which the homogeneous isotropic solution is replaced with a
non-isotropic one. The order parameter in this transition is the
density $\rho_\eta$, and the critical exponent is the same as in the
Landau-Ginsburg theory with positive quartic potential.

In order to complete the picture, and to show that (as expected from
the perturbative analysis) the charged scalar condensate is always the
one with the lowest energy, we will now construct condensates of the
transverse scalar and the vector, and show that their energies are
always higher than the one of the charged scalar. When constructing the
transverse scalar ground state we recall that perturbative analysis
suggested that the s-wave is the first excitation of the scalar which
becomes massless. Hence we make a homogeneous (i.e.~only $u$-dependent)
ansatz as follows
\begin{equation}
\label{fullsolnscal}
A = A_0^{(3)}(u)\tau^3\,{\rm d}t\, , \quad \Phi = \Phi^{(1)}(u)\tau^1 \,, 
\end{equation}
where the vector $A_0$ is present to account for the non-vanishing
chemical potential, while all other vector components are zero.  The
equations of motion for the fields $(A_0, \Phi)$ are given by
\begin{equation}
\label{fulleqnscl}
\begin{aligned}
\partial_z^2A_0^{(3)}(z)+\frac{1+8z^2+3z^4}{z^5-z}\partial_z A_0^{(3)}(z)-\frac{R^2}{z^2}\Phi(z)^2A_0^{(3)}(z) &= 0 \,, \\
\partial_z^2\Phi(z)+\frac{3+4z^2+5z^4}{z^5-z}\partial_z\Phi(z)+\left(\frac{3}{z^2}+\frac{4R^2}{(1+z^2)^2}A_0^{(3)}(z)^2\right)\Phi(z) &= 0 \, .
\end{aligned}
\end{equation}

\begin{figure}[t]
\begin{center}
\includegraphics[width=.48\textwidth]{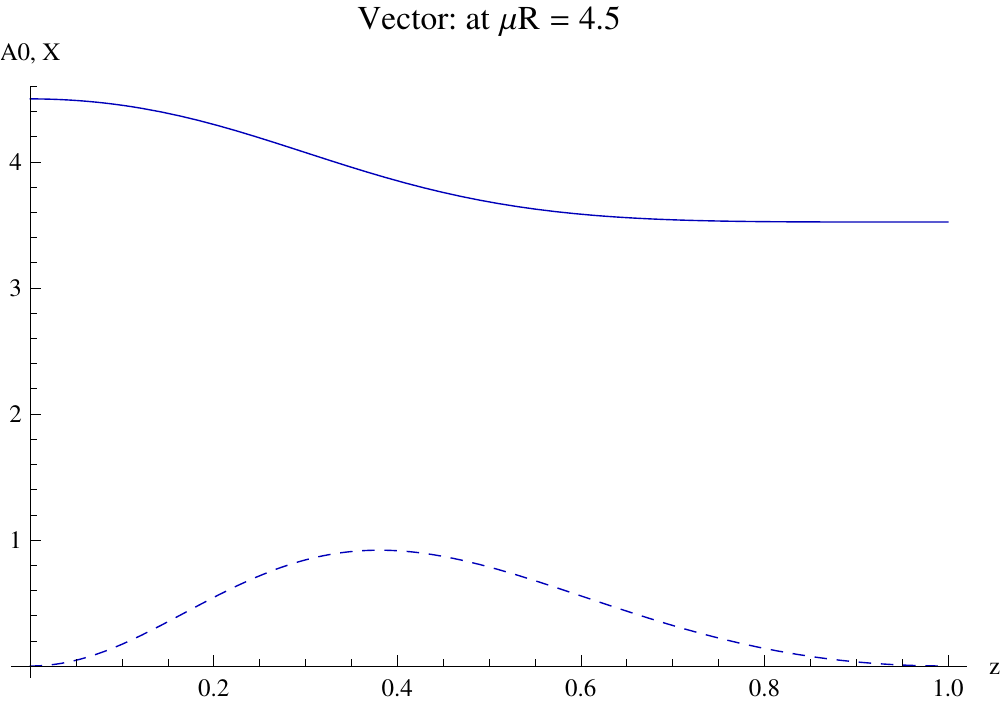}
\includegraphics[width=.48\textwidth]{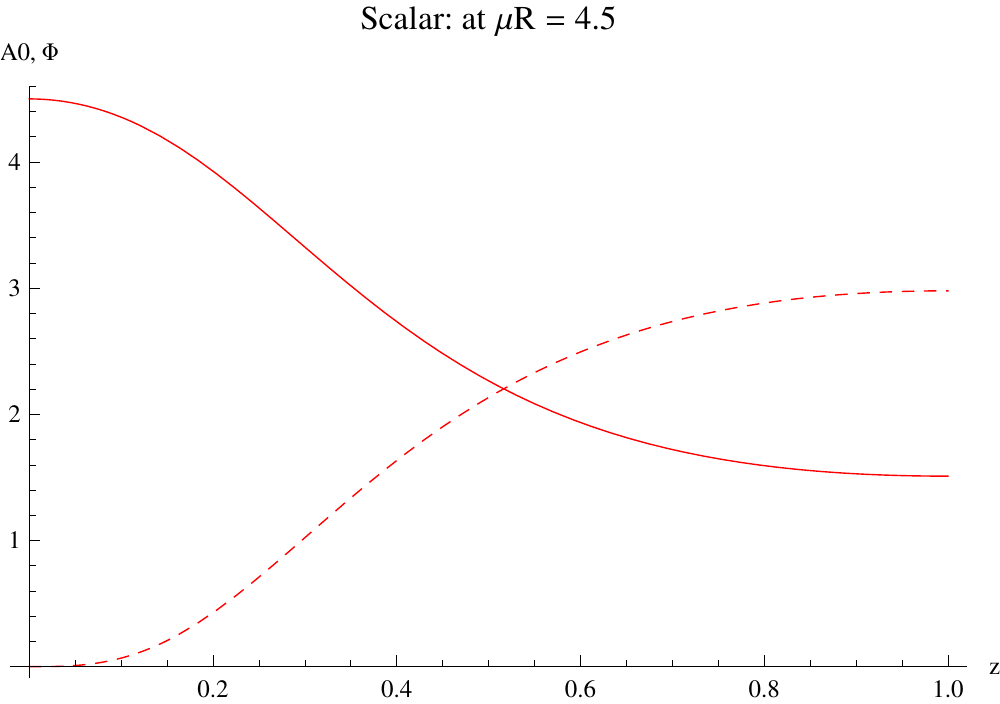}
\end{center}
\caption{Left: a plot of the functions $A_0(z)$ (solid curve) and
  $\psi(z)$ (dashed curve) for the vector solution. Right: plots
  of the functions $A_0(z)$ (solid) and $\Phi$ (dashed) for the scalar
  configuration, both evaluated at a fixed value of chemical potential
  $\mu R =4.5$.
\label{f:solutionscalvect}}
\end{figure}

Similarly, when constructing the ground state originating from vector
condensation, we start with an ansatz which is similar to that of
the vector component dual to a charged scalar, i.e.~we write
\begin{equation}
\label{fullsolutionvect}
A_0 = A_0^{(3)}(u) \tau^3\,,\qquad
A_{\bar{\alpha}} = \psi(u) Y_{\bar{\alpha}}(\bar{\Omega}_3) \tau^1\,,
\end{equation}
where $Y_{\bar{\alpha}}$ is as in \eqref{fullsolution}, except that
the index $\bar{\alpha} = 1,2,3$ now refers to the $\bar{S}^3\in
\text{AdS}_5$. The equations of motion then become
\begin{equation}
\label{fulleqnvecscl}
\begin{aligned}
\partial_z^2A_0^{(3)}(z)+\frac{1+8z^2+3z^4}{z^5-z}\partial_zA_0^{(3)}(z)-\frac{4}{(1-z^2)^2} \psi(z)^2A_0^{(3)}(z) &= 0\,, \\[1ex]
\partial_z^2 \psi(z)+\frac{1+3z^4}{z^5-z}\partial_z \psi(z)+\left(\frac{4R^2}{(1+z^2)^2}A_0^{(3)}(z)^2-\frac{16}{(1-z^2)^2}\right)\psi(z) &= 0 \,.
\end{aligned}
\end{equation}
We should emphasise here that this equation is derived from an ansatz
which uses spherical harmonics with $\bar{l}=1, s=-1$, similar to the
ansatz we used when we constructed the state for the vector dual to a
charged scalar. However, we have also seen in the perturbative
analysis that vector fluctuations in the direction of $\bar{S}^3 \in
\text{AdS}_5$ are insensitive to the quantum number $s$, unlike the
fluctuation in direction of $S^3\in S^5$. Therefore, it should also be
possible to construct an alternative state with spherical harmonics
with $\bar{l}=1, s=1$. This is indeed this the case, and the free
energy of this state is the same as for the state with
$\bar{l}=1,s=-1$.

\begin{figure}[t]
\begin{center}
\includegraphics[width=.48\textwidth]{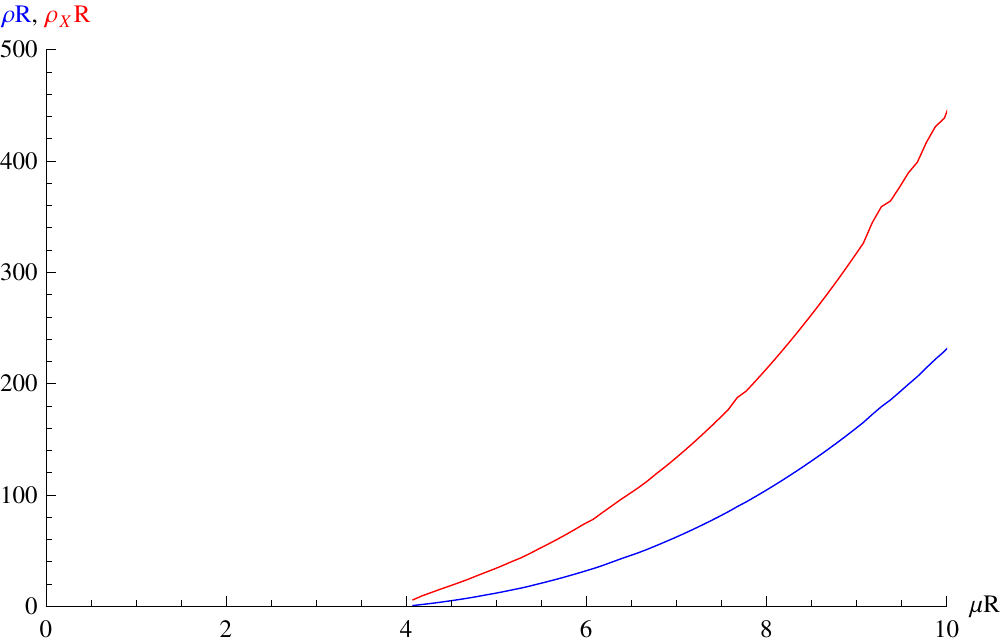}
\includegraphics[width=.48\textwidth]{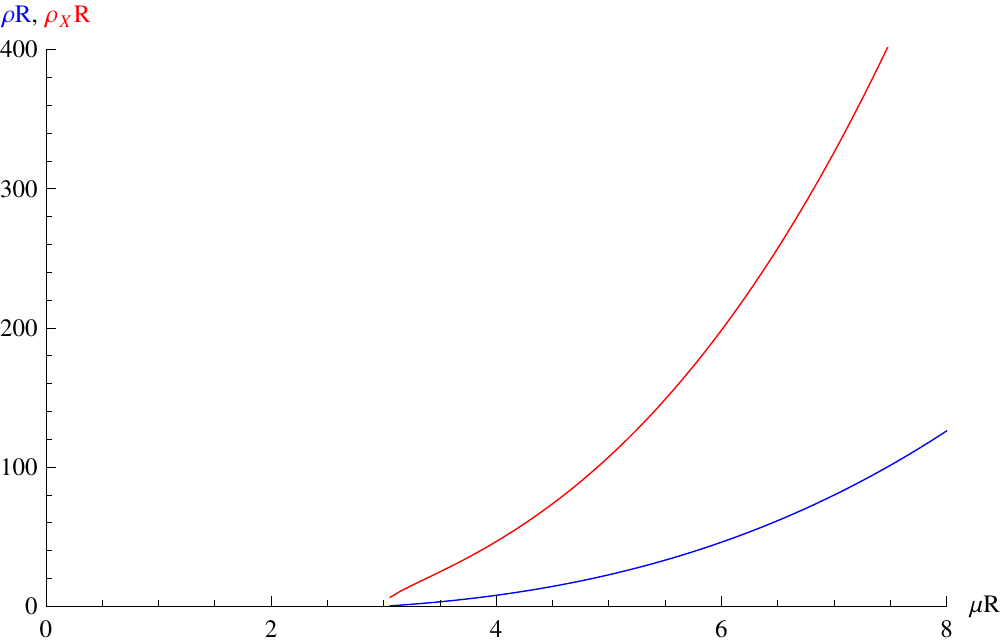}
\end{center}
\caption{Plots of the densities $\rho$ (blue) and $\rho_\eta$ (red),
  as a function of the chemical potential $\mu$, at zero temperature,
  for the vector condensate (left) and scalar condensate (right).
  \label{f:solutionsrhos}}
\end{figure}

Equations (\ref{fulleqnscl}) and (\ref{fulleqnvecscl}) are solved in
the same fashion as equation (\ref{fulleqnvec}), that is by the
shooting method and imposing that the solution is regular everywhere
and in particular at the origin of $\text{AdS}_5$. The solutions for the radial
functions $(A_0,\Phi)$ for the scalar configuration and $(A_0,\psi)$ for
the vector are plotted in figure \ref{f:solutionscalvect}.
We also plot the densities for both configurations (defined analogous
to~\eqref{e:densities_definition}) as functions of the chemical
potential, see figure \ref{f:solutionsrhos}.

\begin{figure}[t]
\begin{center}
\label{solutionrhofnmu}
\vspace{.5cm}
\includegraphics[width=.6\textwidth]{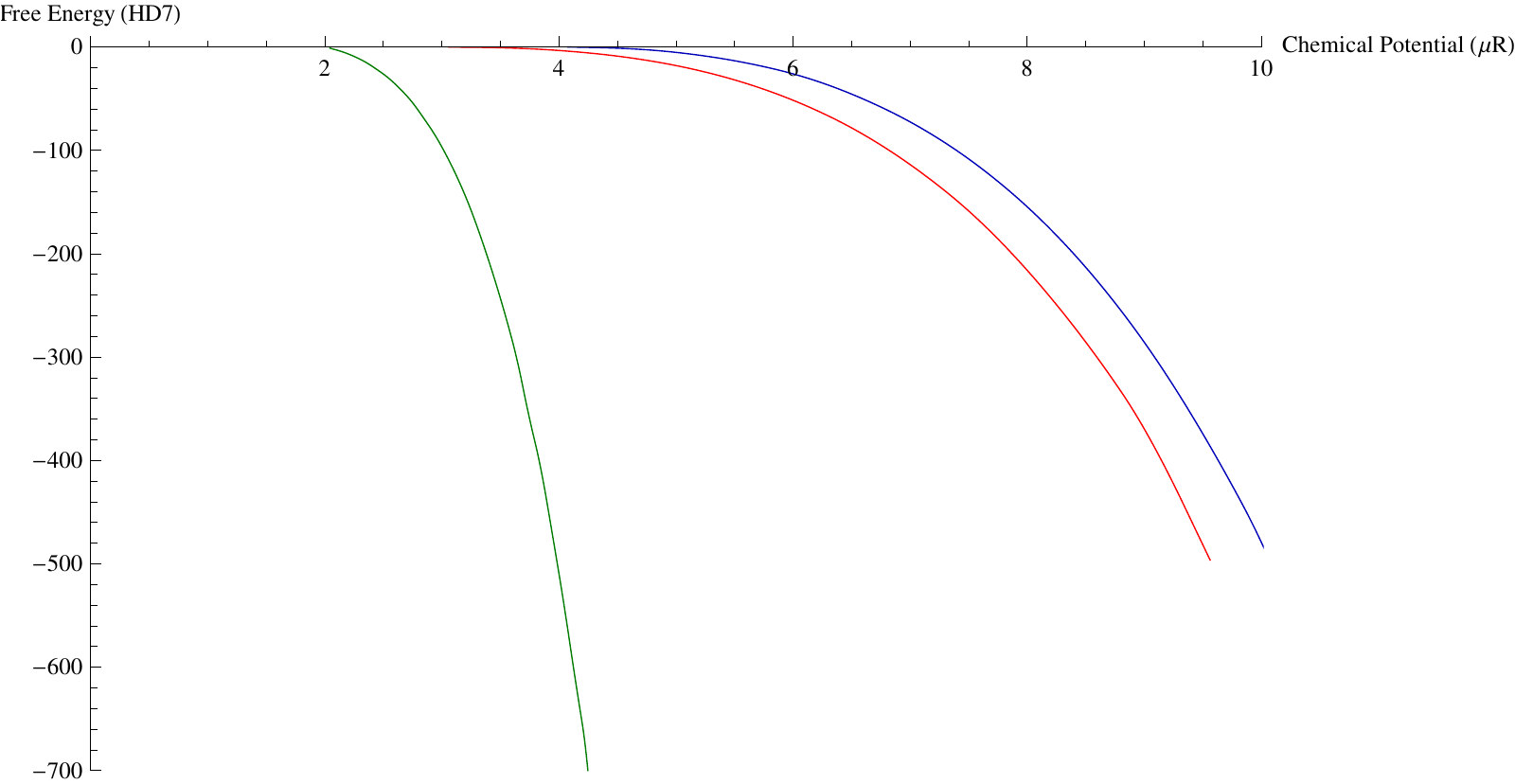}
\vspace{-.5cm}
\end{center}
\caption{Scaled free energy of the zero-temperature condensates as a
  function of the dimensionless chemical potential.  The vector is
  plotted in blue, the scalar in red, the charged scalar green. The
  black line along the $x$-axis denotes the old ground
  state. \label{f:compare_free_energies}}
\end{figure}

In order to compare various configurations we plot the free energies for  all
three states, see figure \ref{f:compare_free_energies}.
As expected from the perturbative analysis, we see that the state
which originates from a condensation of the vector components which
are dual to a charged scalar has the lowest free energy. We also see that as
the chemical potential is increased, the difference between the free
energies of the other two states and the true ground state becomes
larger. It is, however, likely that new instabilities will kick in at
some point. Investigating that in detail would require at the least a
perturbative analysis around this new ground state, which we will not
attempt here.
%\todo{MZ: symmetries of the new state?}

\subsection{The new ground state at finite temperature}

\begin{figure}[t]
\begin{center}
\includegraphics[width=.48\textwidth]{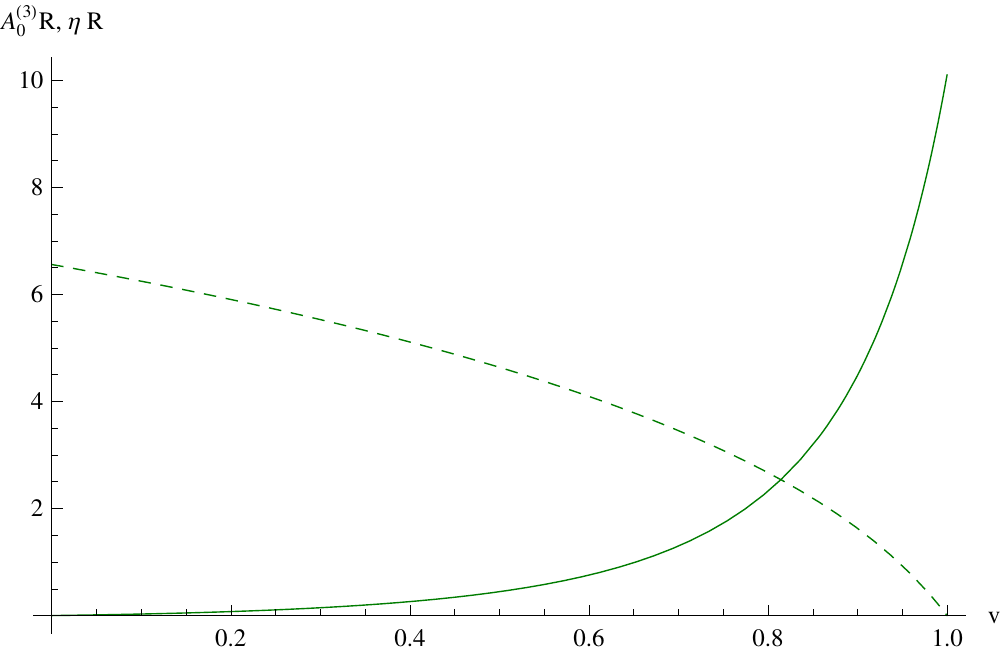} \quad
\includegraphics[width=.48\textwidth]{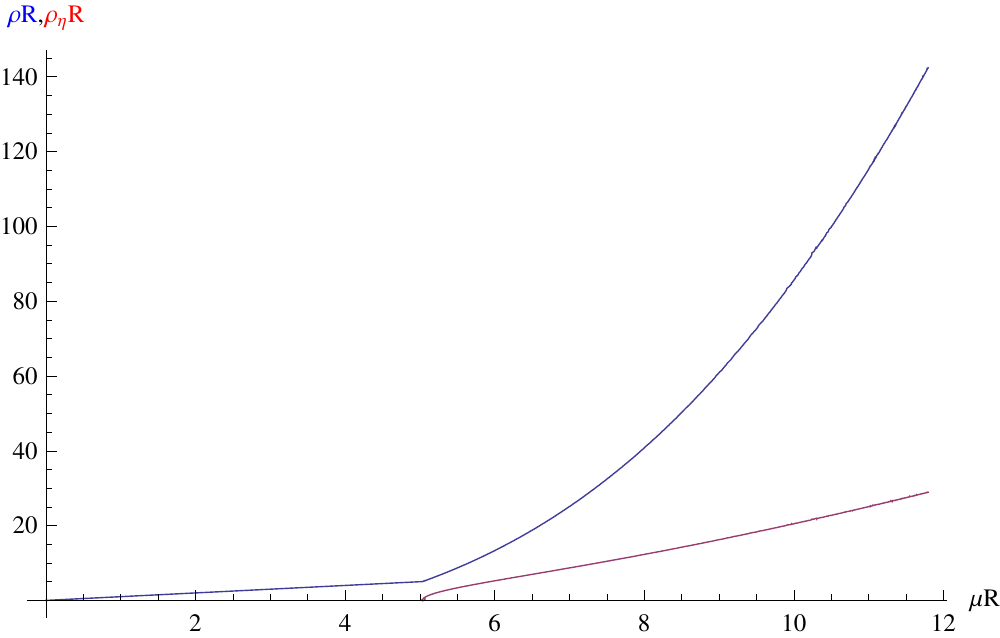}
\end{center}
\caption{Left are plots of profile of the fields $A_0$ (solid)  and
  $\eta$ (dashed) for the charged scalar, evaluated at $\pi TR=2.5$ and $\mu R
  =10.1$. The boundary is at $v=1$ and horizon at $v=0$. Right plot is for densities $\rho$ (blue) and $\rho_\eta$ (red), as function of chemical potential $\mu$, at  fixed temperature $\pi TR=2.5$.
\label{solutionA0etaT}
}
\end{figure}

So far we have seen that at zero temperature, the ground state
originates from the condensation of vector components which are dual
to a charged scalar, exactly as perturbation theory suggested. We now want to
see what is happening with this new ground state as the temperature is
turned on. We start by making the same ansatz as at zero temperature,
see (\ref{fullsolution}). The equations of motion in the coordinates
(\ref{coordinatev}) are given by
\begin{equation}
 \partial_v^2A_0^{(3)}(v)-\frac{\Lambda-1}{4(\Lambda-v)(1-v)^2v}R^2 \eta(v)^2A_0^{(3)}(v)=0\,, 
\end{equation}
together with
\begin{multline}
\partial_v^2\eta(v) +\left(\frac{1}{1-v}+\frac{1}{v}-\frac{1}{\Lambda-v}\right)\partial_v\eta(v)  \\[1ex]
+\frac{\Lambda-1}{(\Lambda-v)(1-v)^2v}\left(1+\frac{(2-\Lambda)(1-v)R^2}{4(\Lambda-v)v}A_0^{(3)}(v)^2\right)\eta(v)
=0\,.
\end{multline}
We are interested in finding regular solutions to these equations. It
is easy to see that the solutions which are regular are parametrised by
two free parameters. A general, perturbative expansion of the solution
near the black hole horizon which is regular is given by
\begin{equation}
\begin{aligned}
A_0^{(3)}(v) &= a v+\frac{a b^2(\Lambda-1)}{8\Lambda R^2}v^2+O(v^3) \,, \\[1ex]
 \eta(v) &= b-\frac{(\Lambda-1)b}{\Lambda}v-\frac{12b(\Lambda-1)+(2-\Lambda)(\Lambda-1)a^2R}{16\Lambda^2}v^2+O(v^3) \,,
\end{aligned}
\end{equation}
i.e.~a regular solution is parametrised by two real numbers $a,b$. We
also see that the general regular solution for $A_0$ vanishes at the
horizon, as required by global regularity.  We solve this system
of equations again using a shooting method with two free parameters. As
at zero temperature, we require in addition that the solution for $\eta$
is normalisable at infinity, or explicitly
\begin{equation}
A_0^{(3)}(v)=\mu-\rho (1-v)+O((1-v)^2)\,,
\quad \eta(v)=\rho_\eta (1-v)+O((1-v)^2)\,.
\end{equation}
This is possible only for a particular pair of parameters $a,b$, or in
other words both densities $\rho, \rho_\eta$ are functions of the
chemical potential. An example of the radial profiles for a regular
solution is plotted in figure \ref{solutionA0etaT}.
We also plot both densities as a function of chemical potential (see
right plot on figure \ref{solutionA0etaT}). We observe that, just as at zero
temperature, the densities increase as the chemical potential is increased.
%\begin{figure}[t]
%\begin{center}
%\label{solutionrhofnmu}
%\includegraphics[width=.7\textwidth]{DSGT0density} 
%\includegraphics[width=.48\textwidth]{DSGT0energy}
%\end{center}
%\caption{ Plots of densities as a function of isospin chemical potential. Blue is isospin density $\rho R.$  and is  density $\rho_{\eta} R.$}
%\end{figure}

\begin{figure}[t]
\begin{center}
\includegraphics[width=.48\textwidth]{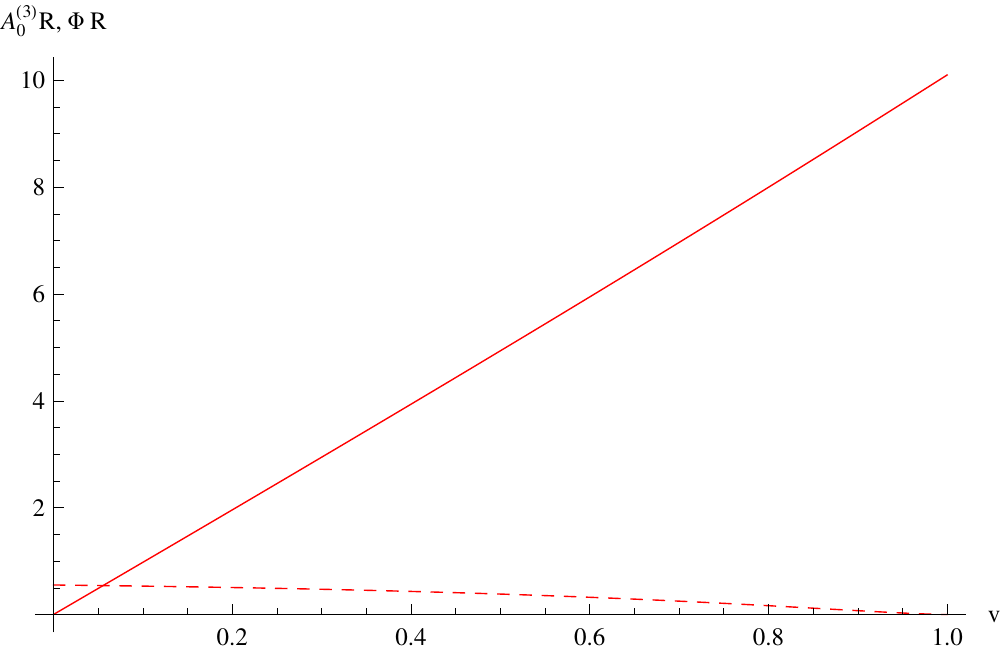}
\includegraphics[width=.48\textwidth]{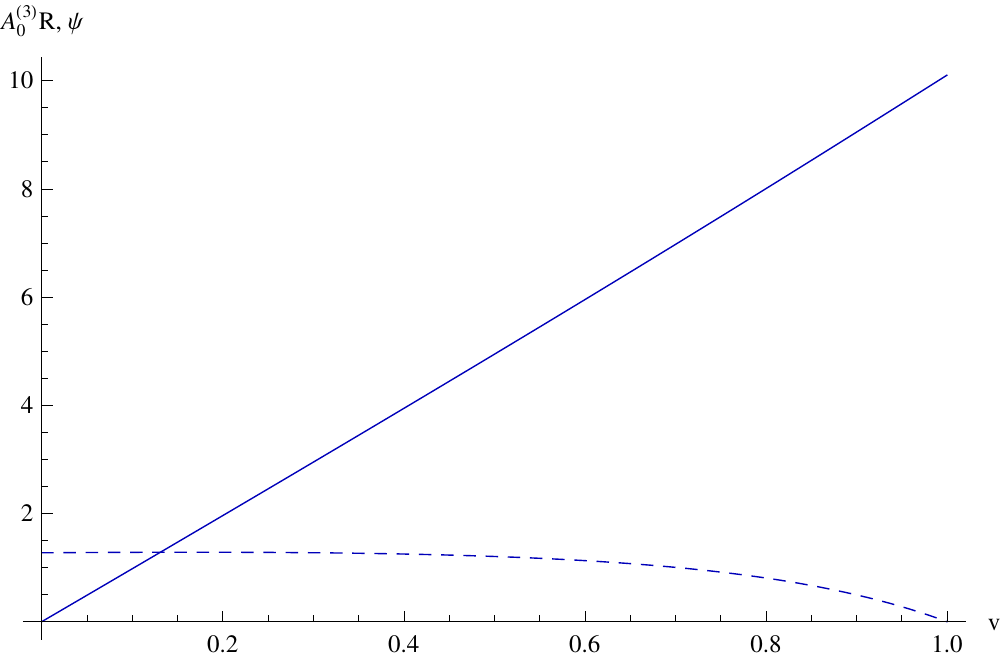}
\end{center}
\caption{Plots of solutions $(A_0(z), \Phi(z))$, and
  $(A_0(z),\psi(z))$ for the scalar and vector condensates
  respectively, at fixed temperature \mbox{$\pi TR=2.5$} and chemical
  potential $\mu R = 4.5$ (the curve for $A_0$ is rather straight
  only because the plot is made for a chemical potential only slightly
  above the critical value). \label{Solscalvect} }
\end{figure}

As for zero temperature, we should make sure that possible alternative
states which appear due to condensation of other unstable particles
have a larger free energy (as suggested by perturbation theory). We start
with the scalar ground state.  We make the same ansatz for the ground
state, as we did at the zero temperature, see
(\ref{fullsolnscal}). The equations of motion in the coordinates
(\ref{coordinatev}) are given by
\begin{equation}
\label{fulleomBHscal}
\begin{aligned}
\partial_v^2A_0^{(3)}(v)&-\frac{\Lambda-1}{4(\Lambda-v)(1-v)^2v}\chi(v)^2
R^2 A_0^{(3)}(v) = 0\,, \\[1ex]
\partial_v^2\chi(v)&+\left(\frac{1}{1-v}+\frac{1}{v}-\frac{1}{\Lambda-v}\right)\partial_v\chi(v) \\[1ex]
	&+\frac{\Lambda-1}{(\Lambda-v)(1-v)^2v}\left(1+\frac{(2-\Lambda)(1-v)R^2}{4(\Lambda-v)v}A_0^{(3)}(v)^2\right)\chi(v) = 0  \, .
\end{aligned}
\end{equation}
Similarly, the ground state originating from the vectors is derived
starting with the ansatz (\ref{fullsolutionvect}). The equations of motion
are given by
\begin{equation}
\label{fulleomBHvect}
\begin{aligned}
\partial_v^2A_0^{(3)}(v)&-\frac{2-\Lambda}{4(\Lambda-v)(1-v)v}\psi(v)^2A_0^{(3)}(v)=0\,, \\[1ex]
\partial_v^2 \psi(v)&+\left(\frac{1}{v}-\frac{1}{\Lambda-v}\right)\partial_v
\psi(v)\\[1ex]
&-\frac{2-\Lambda}{(\Lambda-v)(1-v)v}\left(1-\frac{(\Lambda-1)R^2}{4(\Lambda-v)v}A_0^{(3)}(v)^2\right)\psi(v)=0 \, .  
\end{aligned}
\end{equation}
As before we consider only regular solutions to the equations
(\ref{fulleomBHscal}) and (\ref{fulleomBHvect}) and require that the
solutions are normalisable. Sample solutions for rather arbitrary
values of the temperature and chemical potential are plotted in figure
\ref{Solscalvect}.
We have also evaluated the densities for these solutions, and find a
qualitatively similar dependence on the chemical potential as
before. We have also verified that indeed the charge scalar always has a
lower free energy than the condensate of the other particles, as
predicted by perturbation theory.

\begin{figure}[t]
\begin{center}
\vspace{.5cm}
\includegraphics[width=.6\textwidth]{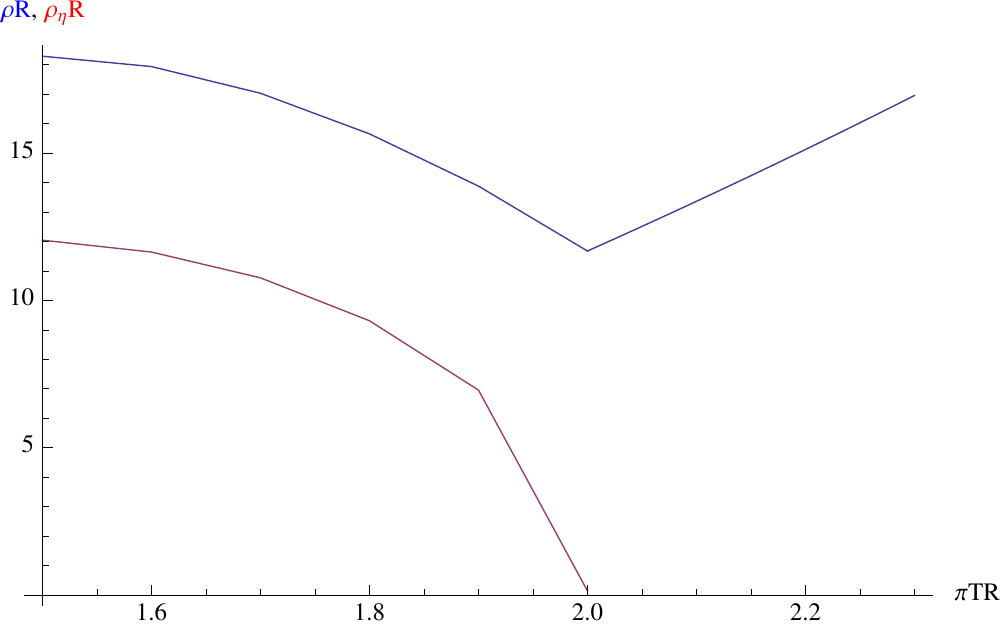}
\vspace{-.5cm}
\end{center}
\caption{Dependence of the charged scalar ground state densities on
  $TR$ at fixed chemical potential $\mu R = 4.005$.\label{f:tempdep}}
\end{figure}

Finally, we present in figure~\ref{f:tempdep} the dependence of the
charged scalar condensate densities on the temperature, for fixed
chemical potential. This shows how increasing the temperature `melts'
the condensate.

\vfill\eject

\section{Discussion and outlook}

In this paper we have analysed the stability of the homogeneous
isotropic phase of conformal ${\mathcal N}=2$ super-Yang-Mills theory on a three
sphere in the presence of an isospin chemical potential. We have found
that for sufficiently large chemical potential, the theory exhibits
unstable vector and scalar modes, as well an unstable vector mode
which is dual to an $SO(4)$ charged scalar in the dual gauge
theory. The latter modes turn out to condense first. We have verified
this explicitly by constructing the condensate and showing that it has
the lowest energy. The new ground state is anisotropic in the
directions of the internal three sphere within the five sphere, but
isotropic on the gauge theory sphere. Therefore, the new ground state
breaks the global $SO(4)$ symmetry.

Since the anisotropy of the system originates from the compactness of
the internal three sphere, this anisotropy persists if we take the limit
towards a non-compact system, i.e.~the limit in which the radius of the
sphere on which the dual gauge theory lives is taken to be very large
(at fixed temperature). 

We have observed that the spectrum of fluctuations of the more massive
vector and scalar mesons crosses as a function of $TR$. This does not
influence the formation of the dual charged scalar condensate, but it
conceivably plays a role for larger values of the chemical
potential. In particular, by doing a fluctuation analysis around the
dual charged scalar condensate, one expects that at some point new
instabilities will set in, corresponding to the more massive particles
condensing as well. An analysis of this type has to our knowledge not
appeared for any string/gauge theory model, but our results indicate
that the structure of condensate formation may be more intricate than
previously suspected.

\acknowledgments

SC is supported by a DPST Scholarship from the Royal Thai
Government. PV is supported by a Durham Doctoral Studentship and by a
DPST Scholarship from the Royal Thai Government.  MZ is supported by
an EPSRC Leadership Fellowship. This work was also partially supported
by STFC grant ST/G000433/1.

%\bibliographystyle{kasper}
%\bibliography{kasbib}

\begingroup\raggedright\endgroup

\end{document}